\documentclass[5p]{elsarticle} 

\usepackage{microtype, textcomp, xspace}
\usepackage[charter, sfscaled, ttscaled, greekuppercase=italicized]{mathdesign}
\usepackage{charter} 
\usepackage{amsmath, physics, isomath, mathtools} 
\usepackage[abbreviations=false, detect-all, list-units=single, math-micro=\muup, text-micro=$\muup$]{siunitx}
\usepackage{graphicx}
\usepackage{booktabs}
\usepackage{makecell, multirow}
\usepackage{url}
\usepackage{parskip}

\usepackage{xcolor}
\colorlet{backgroundcolor}{lightgray!3} 

\usepackage[export]{adjustbox}

\usepackage{listings}

\definecolor{YlGn-K}{RGB}{0,104,55}
\definecolor{GnBu-J}{RGB}{43,140,190}
\definecolor{OrRd-K}{RGB}{179,0,0}
\definecolor{OrRd-H}{RGB}{239,101,72}

\newcommand{\smuthi}{SMUTHI\xspace} 

\newcommand{\fatE}{\mathbf{E}}
\newcommand{\fatr}{\mathbf{r}}

\newcommand{\lmax}{l_\text{max}}
\newcommand{\mmax}{m_\text{max}}

\newcommand{\fatPsiOut}{\mathbf{\Psi}^{(3)}}
\newcommand{\fatPsiReg}{\mathbf{\Psi}^{(1)}}

\newcommand{\init}{\text{init}}
\newcommand{\scat}{\text{scat}}

\newcommand{\ofr}{\left(\fatr\right)}
\newcommand{\neff}{n_\text{eff}}
\newcommand{\dneff}{\Deltaup n_\text{eff}}
\newcommand{\neffmax}{\smash{\neff^\text{max}}}
\newcommand{\neffimag}{\smash{\neff^\text{imag}}}

\usepackage{hyperref}

\journal{JQSRT}

\begin{document}
	
	\begin{frontmatter}
		
		
		\title{\smuthi: A python package for the simulation of light scattering by multiple particles near or between planar interfaces}
				

		
		\author[hepho]{Amos Egel}
		\author[warsaw]{Krzysztof M. Czajkowski\corref{contrib}}
		\author[kit]{Dominik Theobald\corref{contrib}}
		\author[itmo]{\\Konstantin Ladutenko}
		\author[itmo]{Alexey S. Kuznetsov}
		\author[inrim,lens]{Lorenzo~Pattelli}
		
		\cortext[contrib]{Krzysztof Czajkowski and Dominik Theobald contributed equally to this work.}

		\address[hepho]{Hembach Photonik GmbH, 91126 Rednitzhembach, Germany}
		\address[warsaw]{Faculty of Physics, University of Warsaw, 02-093, Warsaw, Poland}
		\address[kit]{Light Technology Institute (LTI), Karlsruhe Institute of Technology (KIT), 76131 Karlsruhe, Germany}
		\address[itmo]{Department of Physics and Engineering, ITMO University, 197101 St.\ Petersburg, Russia}
		\address[inrim]{Istituto Nazionale di Ricerca Metrologica (INRiM), 10135 Torino, Italy}
		\address[lens]{European Laboratory for Non-linear Spectroscopy (LENS), 50019 Sesto Fiorentino, Italy}
		
		\begin{abstract}
			\smuthi is a python package for the efficient and accurate simulation of electromagnetic scattering by one or multiple wavelength-scale objects in a planarly layered medium.
			The software combines the T-matrix method for individual particle scattering with the scattering matrix formalism for the propagation of the electromagnetic field through the planar interfaces.
			In this article, we briefly introduce the relevant theoretical concepts and present the main features of \smuthi.
			Simulation results obtained for several benchmark configurations are validated against commercial software solutions.
			Owing to the generality of planarly layered geometries and the availability of different particle shapes and light sources, possible applications of \smuthi include the study of discrete random media, meta-surfaces, photonic crystals and glasses, perforated membranes and plasmonic systems, to name a few relevant examples at visible and near-visible wavelengths.
		\end{abstract}
		
		\begin{keyword}
			Scattering \sep Multiple scattering \sep T-Matrix \sep Layered media \sep Software
			
			
		\end{keyword}
		
	\end{frontmatter}
	
	
	\section{Introduction}
	\label{S:1}
	
	The efficient collection, extraction or manipulation of light is often based on the interaction between particles and a supporting substrate or a host layered medium.
	Prominent examples of such applications can be found in the fields of metasurfaces, microscopy, plasmonics, illumination and energy harvesting \cite{jahani2016all, staude2017metamaterial, prieve1999measurement, li2010shell, juan2011plasmon, saxena2009review, brongersma2014light}, as well as in more fundamental research areas including cavity electrodynamics, tailored resonances, Anderson localization, topological photonics or bound states in the continuum \cite{yoshie2004vacuum, kuznetsov2016optically, babicheva2017resonant, riboli2014engineering, khanikaev2013photonic, wu2015scheme, hsu2013observation}.
	
	In many cases, the numerical modeling of such systems poses huge computational challenges, especially when dealing with arrangements of many particles that cannot be described as one small unit cell repeated with infinite periodicity.
	Developing new tools to model these systems is therefore a key element to advance their functionalities, improve their design and deepen our understanding of their complex, collective physical behavior.

	\begin{figure}[h]
    	\includegraphics[width=\columnwidth]{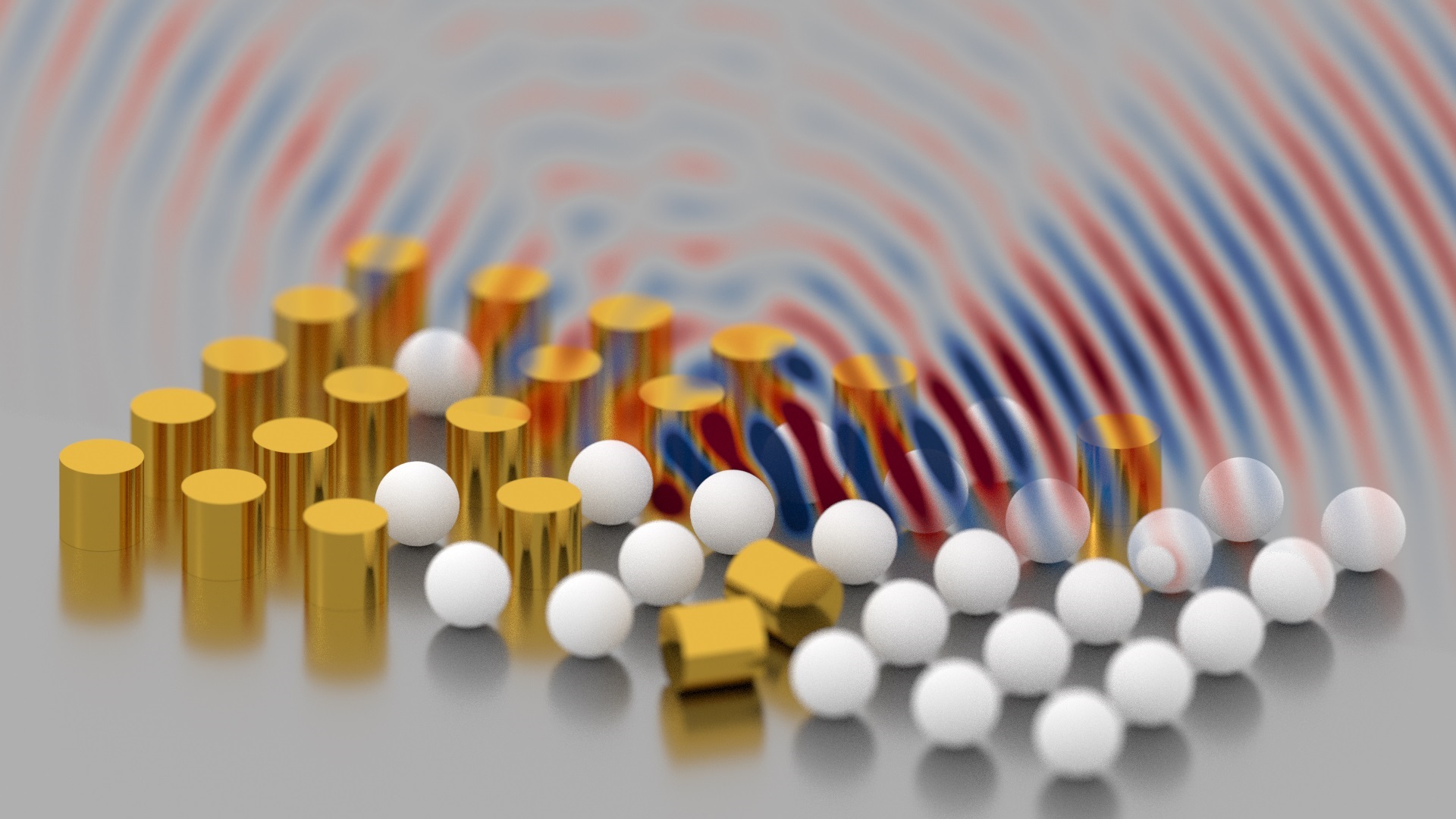}
    	\caption{Artistic visualization of a Gaussian beam scattered by multiple particles on a substrate.}
	\end{figure}

	This paper presents a Python package that allows the optical simulation of light scattering by multiple wavelength-sized particles near or between planar interfaces.
	It implements the superposition T-matrix method \cite{waterman1965matrix, peterson1973t, mackowski1996calculation} which, compared to mesh-based simulation approaches, is characterized by a superior computational efficiency and relatively small memory footprint, thereby enabling the simulation of large systems that would be impossible to model otherwise.

	Several open source implementations dedicated to the simulation of electromagnetic scattering by multiple particles using the superposition T-matrix method have been made available in the last decade \cite{mackowski2011multiple, markkanen2017fast, egel2017celes}.
	Despite their computational efficiency, however, these implementations are limited to the case of particles in a homogeneous background.

	The here presented software lifts this limitation, by combining the advantages of the T-matrix method with the possibility to model systems involving planar interfaces.

	The paper is organized as follows: Section \ref{sec:smuthi} introduces the open source project behind the software.
	In section \ref{sec:method}, a brief overview over the underlying theoretical concepts is given. 
	A short user guide is provided in section \ref{sec:guide}.
	Section \ref{sec:examples} illustrates some typical use case examples and establishes the validity of the code by comparison to accurate benchmark results from third-party software.
	Finally, we present conclusions and outlook in section \ref{sec:conclusions}.
	
	\section{The \smuthi project}
	\label{sec:smuthi}
	\smuthi (``Scattering by MUltiple particles in THIn film systems'') is a Python package published under the MIT license.
	It is an open source project, initiated by the first author of this paper and further developed collaboratively by a community of scientists from different research groups.
	
	The project is hosted on:
	\begin{itemize}
		\item \url{https://gitlab.com/AmosEgel/smuthi} (code repository)
		\item \url{https://pypi.org/project/SMUTHI} (Python package)
		\item \url{https://smuthi.readthedocs.io} (documentation)
		\item \url{https://groups.google.com/g/smuthi} (mailing list)
	\end{itemize}

	The source code follows an object oriented programming style.
	This allows easy access to individual modules of the code, which can be adapted to specific needs of the user or imported into other software projects.
	
	For more demanding simulation tasks, \smuthi allows parallelization of performance-critical operations on an NVIDIA graphics processing unit through the PyCUDA package \cite{klockner2012pycuda}.

	The target group of users are scientists and engineers in the field of optics and electromagnetic scattering.
	Although \smuthi offers some tools to assist the user in selecting appropriate numerical settings, a critical evaluation of the numerical results is always required to verify convergence and accuracy.

	\section{Simulation method}
	\label{sec:method}
	
	\smuthi solves the 3D Maxwell equations in the frequency domain (i.e., one wavelength per simulation).
	It is based on the T-matrix method \cite{waterman1965matrix,mishchenko2004t} which was adapted to treat multiple particles (superposition T-matrix method, \cite{peterson1973t,mackowski1996calculation}) located near or between planar interfaces \cite{kristensson1980electromagnetic,mackowski2008exact,egel2014dipole}.
	
	One important advantage of the T-matrix method for multiple particles is its \emph{modularity}: The evaluation of scattering by a single particle is separated from the evaluation of the mutual interaction between the particles and from the evaluation of the interaction between a particle and a non-homogeneous environment, such as a particle on a substrate.
	Once a particle's T-matrix is known, its scattering behavior can be evaluated under (almost) arbitrary illumination conditions.
	
	For the evaluation of the T-matrix of an isolated particle, \smuthi relies on the state-of-the-art NFM-DS (``Null-field method with discrete sources'') FORTRAN code by Adrian Doicu, Thomas Wriedt and Yuri Eremin \cite{doicu1999calculation,doicu2006light}.
	T-matrices computed by NFM-DS are then processed by \smuthi to solve the problem of multiple scattering between the particles as well as between the particles and the layered medium.
	
	In the following subsection, we sketch the basic concepts to provide a general outline of the method.
	The interested reader is referred to chapters 2-3 of \cite{egel2017accurate} for a more complete description of the underlying theory.
	
	\subsection{The T-Matrix method for multiple particles between planar interfaces}
	
	In the context of scattering particles located near or inside planarly layered media, it is useful to present the total electric field as the sum of four constituent parts:
	\begin{align}
	\fatE\ofr = \fatE_\init\ofr + \fatE_\init^R\ofr + \sum_{i=1}^{N_S} \fatE_{\scat,i}\ofr + \sum_{i=1}^{N_S} \fatE^R_{\scat,i}\ofr,
	\end{align}
	where $\fatE_\init$ denotes the \emph{initial excitation}, $\fatE_\init^R$ the response of the layer system to that field, $\fatE_{\scat,i}$ the scattered field from particle $i$, 
	and  $\fatE_{\scat,i}^R$ the response of the layer system to that field (the sums run over 
	all $N_S$ particles).

	The central aspect of the T-matrix method is the expansion of the electromagnetic field using spherical vector wave functions.
	The incoming field at particle $i$ (i.e., the initial field including layer response as well as the scattered field from all other particles including the layer response) is expanded in terms of regular spherical vector wave functions, $\fatPsiReg_n$ (see section B.2 of \cite{doicu2006light}),
	\begin{spreadlines}{1.5ex}
	\begin{equation}
	\label{eq:expansion2}	
	\left.
	\begin{alignedat}{2}
	\fatE_{\init}\ofr &= \sum_{n} a^i_{\init,n} \fatPsiReg_{n} (\fatr - \fatr_i)
	\\ \fatE_{\init}^R\ofr &= \sum_{n} a^{i,R}_{\init,n} \fatPsiReg_{n} (\fatr - \fatr_i) \;
	\\ \fatE_{\scat,j}\ofr &= \sum_{n} a^i_{j,n} \fatPsiReg_{n} (\fatr - \fatr_i) \;
	\\ \fatE_{\scat,j}^R\ofr &= \sum_{n} a^{i,R}_{j,n} \fatPsiReg_{n} (\fatr - \fatr_i) \;	
	\end{alignedat}
	\right\} \text{ for }\left| \fatr-\fatr_i \right| \leq R^\text{in}_i,
	\end{equation}
	\end{spreadlines}
	where $R^\text{in}_i$ is radius of the largest sphere fully contained inside particle $i$.

	On the other hand, the scattered electromagnetic field is expanded in terms of outgoing spherical vector wave functions, $\fatPsiOut_{n}$ (see section B.2 of \cite{doicu2006light}),
	\begin{align}
	\label{eq:expansion1}
	\fatE_{\scat,i}\ofr &= \sum_{n} b^i_{n} \fatPsiOut_{n} (\fatr - \fatr_i) &\text{for }\left| \fatr-\fatr_i \right| \geq R^\text{um}_i,
	\end{align}
	where $b^i_n$ are the expansion coefficients, $\fatr_i$ is the center coordinate of particle $i$, and $R^\text{um}_i$ is the radius of the smallest sphere that contains particle $i$.
	In the above summation, we have applied a multi-index notation, where $n$ subsumes the spherical polarization $\tau=(\text{TE},\text{TM})$, the multipole degree $l=1,\ldots,\infty$, and order $m=-l,\ldots,l$.
	Note that the expansion \eqref{eq:expansion1} is strictly valid only outside the circumscribing sphere of particle $i$.

	In total, we are thus dealing with five sets of expansion coefficients, two of which (the initial field coefficients and the initial field layer response coefficients) are known \emph{a priori} and three of which are unknown.
	Correspondingly, there are three sets of linear equations that can be used to determine the unknown expansion coefficients.
	First, the T-matrix equation connects the incoming to the scattered field coefficients for each particle,
	\begin{align}
	\label{eq:T}
	b_n^i = \sum_{n'}T^i_{nn'}\left(a_{\init,n'}^{i} + a_{\init,n'}^{i, R} + \sum_{j\neq i} a_{j,n'}^{i} + \sum_{j\neq i} a_{j,n'}^{i, R} \right),
	\end{align}	
	where $T^i$ denotes the T-matrix of particle $i$.
	Second and third, the coupling matrices connect the scattered field coefficients of particle $j$ to the corresponding incoming field coefficients of particle $i$:
	\begin{align}
	\label{eq:aj}	
	a_{j,n}^i &= \sum_{n'}W^{i}_{j,nn'} b_{n'}^{j}
	\\
	\label{eq:ajR}
	a_{j,n}^{i,R} &= \sum_{n'}W^{i,R}_{j,nn'} b_{n'}^{j}
	\end{align}	
	The above coupling equations define the direct ($\smash{W^{i}_{j}}$) and layer system mediated ($\smash{W^{i,R}_{j}}$) particle coupling operator.
	
	The direct coupling operator describes how an outgoing spherical wave, emitted from position $\fatr_j$, will be perceived as a series of regular spherical waves at position $\fatr_i$.
	This translation relation is given by the spherical vector wave function addition theorem \cite{stein1961addition,cruzan1962translational}, which can either be constructed from closed form expressions involving Wigner-3$j$ functions \cite{mishchenko2002scattering, johansson2016fast} or from an iterative scheme \cite{doicu2006light}.
	
	The layer system mediated coupling operator, on the other hand, states how the multiple transmissions and reflections of an outgoing spherical wave, emitted from position $\fatr_j$ and propagated through the layer system, are then perceived as a series of regular spherical waves at position $\fatr_i$.
	Note that, due to the lateral translation symmetry of the planarly layered medium, the calculation of the layer system mediated coupling operator is most conveniently done using plane wave expansions rather than spherical waves.
	It is constructed by the following procedure:
	\begin{enumerate}
		\item Expand the outgoing spherical waves at $\fatr_j$ into plane waves.
		\item Propagate the plane waves through the layer system (using the transfer matrix method or the scattering matrix method, \cite{egel2014dipole,egel2019accurate}).
		\item Expand each plane wave back into regular spherical waves at $\fatr_i$.
	\end{enumerate}
	Further details are beyond the scope of this work and can be found in \cite{egel2019accurate}.
	The resulting expression for $\smash{W^{i,R}_j}$ (see equation (3.46) of \cite{egel2019accurate}) involves an integral (the so-called \emph{Sommerfeld integral}), which in general has to be solved numerically.
	
	Finally, equations \eqref{eq:aj} to \eqref{eq:ajR} can be inserted into \eqref{eq:T} to yield the following linear system
	\begin{align}
	\label{eq:linear system}
	\sum_{j\neq i} \sum_{n'} M^i_{j,nn'} b_{n'}^i = \sum_{n'}T_{nn'}^i \left( a_{\init,n'}^i + a_{\init,n'}^{i,R} \right)
	\end{align}	
	with
	\begin{align}
	\label{eq:master}
	M^i_{j,nn'} = \delta_{ij}\delta_{nn'} - \sum_{n''} T^i_{nn''} \left( W^i_{j,n''n'} + W^{i,R}_{j,n''n'} \right).
	\end{align}	
	
	Solving \eqref{eq:linear system} yields the scattered field coefficients, from which all quantities of interest can then be computed in a post-processing step.
	
	It is worth noting that the dimension of the linear system \eqref{eq:linear system} is proportional to the number of particles and to the number of partial waves used in the multipole expansions \eqref{eq:expansion1}.
	For systems with many particles, solving \eqref{eq:linear system} can quickly become a numerically substantial task.
	
	\section{User guide}
	\label{sec:guide}	
	This section provides an overview of the basic principles that govern a \smuthi simulation and allows a new user to get started with the simulation of standard application scenarios.
	
	The following instructions were tested for \smuthi version 1.2.4.
	Possible future changes will be documented on the online documentation, to which the user is referred for a full description of \smuthi application programming interface (API).
	
	\subsection{Installation}
	
	\subsubsection{Hardware and software requirements}	
	For simple simulations (e.g., a single particle on a substrate), \smuthi does not have special hardware requirements.
	For heavier simulations comprising a large number of particles or involving demanding post processing steps, we recommend the use of a workstation computer with sufficient memory and with a CUDA-capable NVIDIA graphics card.
	Alternatively, a straightforward way to test \smuthi is offered by online Jupyter platforms such as Google Colab \cite{colab}, which also provide free access to NVIDIA GPU hardware.
	
	In order to run \smuthi, Python 3.6 (or later) and the ``Pip'' Python package manager are required.
	Assuming a Linux (e.g., Ubuntu) operating system, the Foreign Function Interface library may be also needed, which can be installed with 
	
	\colorbox{gray}{\color{white} \rlap{\texttt{sudo apt install libffi6 libffi-dev}}\hspace{\linewidth}\hspace{-2\fboxsep}}
	
	In order to benefit from CUDA-accelerated calculations, a suitable NVIDIA graphics card, the NVIDIA CUDA toolkit and the PyCuda Python package are also needed.
	To install the latter, run
	
	\colorbox{gray}{\color{white} \rlap{\texttt{sudo python3 -m pip install pycuda}} \hspace{\linewidth}\hspace{-2\fboxsep}}
	
	\subsubsection{Installation}	
	The following command installs the latest \smuthi release from the Python Package Index (PyPi):
	
	\colorbox{gray}{\color{white} \rlap{\texttt{sudo python3 -m pip install smuthi}} \hspace{\linewidth}\hspace{-2\fboxsep}}
	
	Alternatively, download the \smuthi sources from the online Git repository (\url{https://gitlab.com/AmosEgel/smuthi}), browse into the project folder and install it manually by
	
	\colorbox{gray}{\color{white} \rlap{\texttt{sudo python3 -m pip install .}} \hspace{\linewidth}\hspace{-2\fboxsep}}

	\subsection{``Hello World''}
	
	The following code represents a minimal simulation example -- a ``Hello World'' \smuthi script.
	It simulates the extinction cross section for a glass sphere on a glass substrate:
	
	\begin{lstlisting}
import numpy as np
from smuthi.simulation import Simulation
from smuthi.initial_field import PlaneWave
from smuthi.layers import LayerSystem
from smuthi.particles import Sphere
from smuthi.postprocessing.far_field \
     import extinction_cross_section

laysys = LayerSystem(thicknesses=[@0@,@0@],
                     refractive_indices=[@1.52@,@1@])

sph = Sphere(position=[@0@,@0@,@100@],
             refractive_index=@1.52@,
             radius=@100@,
             l_max=@3@)

plwv = PlaneWave(vacuum_wavelength=@550@,
                 polar_angle=np.pi,
                 azimuthal_angle=@0@,
                 polarization=@0@)

simul = Simulation(layer_system=laysys,
                   particle_list=[sph],
                   initial_field=plwv)
simul.run()

ecs = extinction_cross_section(simulation=simul)
print("Extinction cross section:", ecs)
	\end{lstlisting}	
	
	\subsection{Building blocks of a simulation}	
\begin{figure}[ht]
	\centering
	\resizebox{\columnwidth}{!}{\includegraphics[width=\columnwidth]{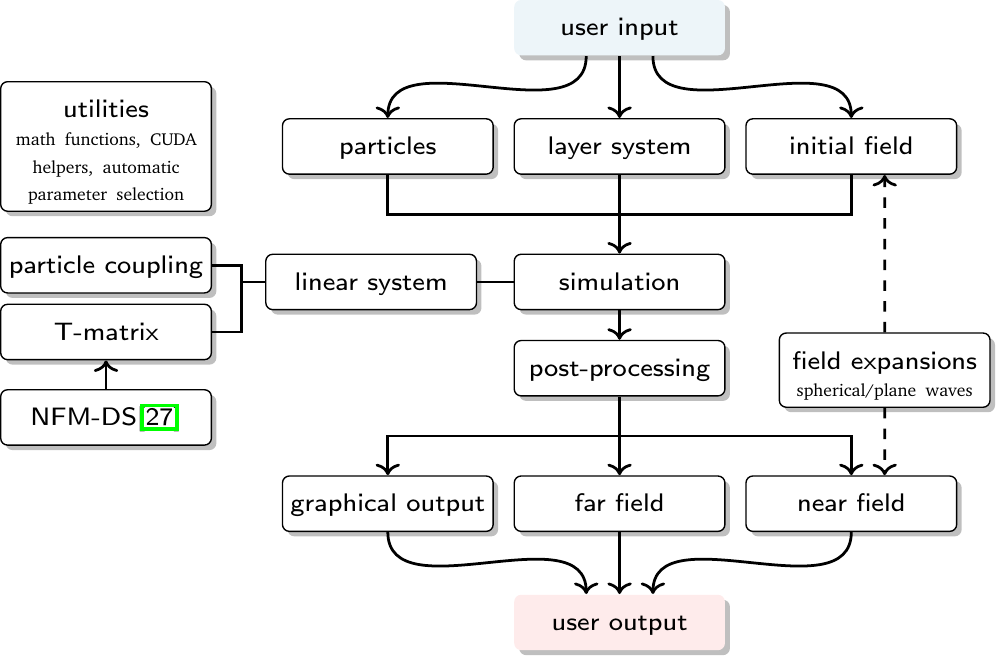}}
	\caption{\smuthi module structure and data flow}
	\label{fig:modules}
	\end{figure}

	In general, a \smuthi simulation script contains the following building blocks (compare figure \ref{fig:modules}):
	
	\begin{itemize}
		\item Definition of the optical system: the initial field, the layer system and a list of scattering particles are defined
		\item Definition of the simulation object: the \lstinline!simulation! object is initialized with the ingredients of the optical system.
			  Further numerical settings can be applied.
		\item Simulation: The calculation is launched with the command \lstinline!simulation.run()!
		\item Post processing: The results are processed into the desired output (for example: scattering cross section).
	\end{itemize}

	The following sections contain a description of each of these building blocks, together with some example code snippets that illustrate their usage in a \smuthi simulation script.
	
	\subsubsection{Initial field}
	Currently, the following classes can be used to define the initial field:
	
	\paragraph{Plane waves}
	An initial plane wave is specified by the vacuum wavelength, incident direction, polarization, complex amplitude and reference point.
	For example:

\begin{lstlisting}
import numpy as np
from smuthi.initial_field import PlaneWave

plwv = PlaneWave(vacuum_wavelength=@550@,
                 polar_angle=np.pi,
                 azimuthal_angle=@0@,
                 polarization=@0@)   # 0=TE 1=TM
\end{lstlisting}
	
	\paragraph{Gaussian beams}
	A Gaussian beam is specified by the vacuum wavelength, incident direction, polarization, complex amplitude, beam waist and reference point.
	Note that for oblique incident directions, the Gaussian beam is in fact an elliptical beam, such that the electric field in the $xy$-plane, i.e., parallel to the layer interfaces has a circular Gaussian footprint.
	For details, see \cite{egel2019accurate}.
	
	A typical Gaussian beam can be implemented as follows.
\begin{lstlisting}
import numpy as np
from smuthi.initial_field import GaussianBeam

focus = [@0@, @0@, @500@]
beam = GaussianBeam(vacuum_wavelength=@550@,
                    polar_angle=np.pi,
                    azimuthal_angle=@0@,
                    polarization=@0@,
                    reference_point=focus,
                    beam_waist=@5000@)
\end{lstlisting}

	Further parameters can be set to control the Gaussian beam numerical precision.
	For details, see the API section of the online documentation.
		
	\paragraph{Point dipoles}
	A single point dipole source is specified by the vacuum wavelength, dipole moment vector and position.
	The following code illustrates the definition of a dipole source:

\begin{lstlisting}
from smuthi.initial_field import DipoleSource

# initialize dipole object
dip = DipoleSource(vacuum_wavelength=@550@,
                   dipole_moment=[@1@,@0@,@0@],
                   position=[@0@,@0@,@300@])
\end{lstlisting}

	Further parameters can be set to control the dipole source numerical precision.
	For details, see the API section of the online documentation.
	
	To define multiple dipole sources, create a dipole collection:
\begin{lstlisting}
from smuthi.initial_field import DipoleSource, \
                                 DipoleCollection

# initialize dipole objects
dip1 = DipoleSource( ... )
dip2 = DipoleSource( ... )
dip3 = DipoleSource( ... )

# add them to a collection
collection = DipoleCollection(vacuum_wavelength=ld)
collection.append(dip1)
collection.append(dip2)
collection.append(dip3)
\end{lstlisting}

	Note: a point dipole source must not be placed inside the circumscribing sphere of a scattering particle, except if the dipole is in a different layer than the particle.
	
	\subsubsection{Layer system}
	
	The layer system is specified by a list of layer thicknesses and a list of complex refractive indices.
	
	Note that:
	\begin{itemize}
		\item The layer system is built from bottom to top, i.e., the first elements in the lists refer to the bottom layer.
		\item The interface between the bottom layer and the next layer in the layer system defines the $z=0$ plane.
		\item Bottom and top layer are semi-infinite in size.
			  You can specify a layer thickness of zero.
		\item The minimal layer system consists of two layers (e.g., a substrate and an ambient medium).
			  Homogeneous media without layer interfaces cannot be defined, but they can be mimicked by a trivial system of two identical layers.
			  However, pure T-matrix implementations without the computational overhead due to planar interfaces should be preferred for this simple case.
	\end{itemize}

	A typical layer system definition could look like this (metal substrate with dielectric coating under vacuum):
\begin{lstlisting}
from smuthi.layers import LayerSystem

# layer refractive indices
n0 = @1.0@+@6.1j@	# bottom, metal
n1 = @1.45@			# middle, coating
n2 = @1@				# top, vacuum

# layer thicknesses
d0 = @0@				# bottom and
d1 = @120@			# top must be
d2 = @0@				# set to d=0

laysys = LayerSystem(thicknesses=[d0,d1,d2],
                     refractive_indices=[n0,n1,n2])
\end{lstlisting}

	\subsubsection{Particles}
	
\begin{figure}
		\centering
		\includegraphics{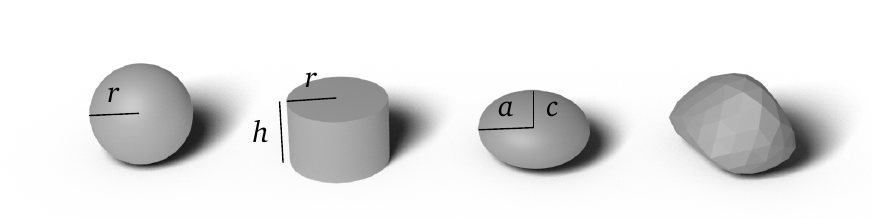}
		\caption{Illustration of possible particle shapes.
				 From left to right: sphere, cylinder, spheroid, custom particle}
				 \label{fig:particles}
	\end{figure}

	\smuthi supports simulations comprising different types and combinations of scattering particles. A few typical geometries are shown in figure \ref{fig:particles}.
	
	\paragraph{Spheres}
	A sphere is specified by its center position vector, complex refractive index, radius and multipole cutoff, e.g.:
	
\begin{lstlisting}
from smuthi.particles import Sphere

sph = Sphere(position=[@0@,@0@,@300@],
             refractive_index=@2.4@,
             radius=@110@,
             l_max=@3@,
             m_max=@3@)
\end{lstlisting}	

	\paragraph{Spheroids}
	A spheroid is specified by its center position vector, Euler angles (to define the orientation in space), complex refractive index, two half axis parameters, and the multipole cutoff, e.g.:

\begin{lstlisting}
from smuthi.particles import Spheroid

sphrd = Spheroid(position=[@0@,@0@,@200@],
                 euler_angles=[@0.3@,@0.5@,@1.2@],
                 refractive_index=@2.4@,
                 semi_axis_c=@200@,
                 semi_axis_a=@100@,
                 l_max=@5@,
                 m_max=@5@)
\end{lstlisting}	
	
	\paragraph{Cylinders}
	A cylinder is specified by its center position vector, Euler angles, complex refractive index, radius, height and multipole cutoff, e.g.:

\begin{lstlisting}
from smuthi.particles import FiniteCylinder

cyl = FiniteCylinder(position=[@0@,@0@,@200@],
                     euler_angles=[@0.3@,@0.5@,@1.2@],
                     refractive_index=@2.4@,
                     cylinder_radius=@80@,
                     cylinder_height=@140@,
                     l_max=@5@,
                     m_max=@5@)
\end{lstlisting}	

	\paragraph{Custom particle}
	The custom particle class allows to model particles with arbitrary geometry.
	These are specified by their position vector, Euler angles, an STL (or, alternatively, FEM) file containing the particle surface mesh, a scale parameter to set the physical size of the particle (if it deviates from the size specified by the mesh file) and multipole cutoff.

\begin{lstlisting}
from smuthi.particles import CustomParticle

cust = CustomParticle(position=[@0@,@0@,@200@],
                      euler_angles=[@0.3@,@0.5@,@1.2@],
                      refractive_index=@2.4@,
                      geometry_filename="custom.stl",
                      scale=@100@,
                      l_max=@5@,
                      m_max=@5@)
\end{lstlisting}	

	Two useful tools to generate an STL mesh file for a given geometry are GMSH \cite{geuzaine2009gmsh} and trimesh \cite{trimesh}.	

	When defining a scattering particle, you need to provide the parameters regarding geometry and material, as well as the parameters \lstinline!l_max! and \lstinline!m_max! which define the multipole expansion cutoff and can be specified for each particle independently, see section \ref{sec:cutoff}.
	
	Additional notes:
	\begin{itemize}
		\item At the moment, the simulation of non-spherical particles relies on the NFM-DS Fortran code by Adrian Doicu, Thomas Wriedt and Yuri Eremin \cite{doicu2006light}.
		\item Particles must not overlap with each other or with layer interfaces.
		\item The circumscribing spheres of non-spherical particles may overlap with layer interfaces (e.g., a flat particle on a substrate), but care has to be taken with regard to the selection of the numerical parameters \cite{egel2016light,egel2017extending}.
			  Use of \smuthi's automatic parameter selection feature is recommended, see section \ref{sec:autoparam}.
		\item The circumscribing spheres of non-spherical particles must not overlap with each other.
			  There is a \smuthi sub-package which offers plane-wave mediated particle coupling and thus allows treating particles with overlapping circumscribing spheres \cite{theobald2017plane}, but this feature is still experimental and requires expert knowledge to be used.
	\end{itemize}

	\subsubsection{The simulation class}
	The \lstinline!Simulation! object is the central manager of a \smuthi simulation.
	To define a \lstinline!Simulation!, you need to at least specify the optical system, i.e., an initial field, a layer system and a list of scattering particles.
	In addition, you can provide a number of input parameters regarding numerical parameters or solver settings, see section \ref{sec:numpar}.
	
	The following code illustrates the definition of a simulation object, using the default settings for numerical parameters:
	
\begin{lstlisting}
from smuthi.simulation import Simulation

# initialize simulation
simul = Simulation(layer_system=laysys,
                   particle_list=[sph, cyl],
                   initial_field=beam)

# run simulation
simul.run()
\end{lstlisting}	

	\subsubsection{Post-processing}
	Once the \lstinline!Simulation.run()! method has successfully terminated (i.e., all unknown expansion coefficients have been determined), we still need to process the results into the desired simulation output.
	\smuthi offers data structures to obtain near and far field distributions as well as scattering cross sections.
	Below, we give a short overview on a couple of convenience functions that can be used to quickly generate some output.
	
	\paragraph{Near fields}
	The near field\footnote{The term ``near field'' is opposed to ``far field'' which is an intensity distribution in direction space.
	It does not imply that the field is evaluated near the particles.} is and electric field distribution as a function of position, $\fatE=\fatE(\fatr)$.
	
	The following code snippets illustrates the generation of electric field plots and animations.
	
\begin{lstlisting}
from smuthi.postprocessing.graphical_output \
     import show_near_field

qts_to_plot = ["norm(E)", "E_y"]

show_options = [{"label":"raw_data"},
                {"interpolation":"quadric"},]
                
show_near_field(simulation=simul,
                quantities_to_plot=qts_to_plot,
                show_opts=show_options,
                xmin=@-600@, xmax=@600@,
                zmin=@-100@, zmax=@900@,
                show_internal_field=@True@)
\end{lstlisting}		
	For a full list of possible settings, see the online documentation.
	Note: Spheres allow the evaluation of near fields everywhere (inside and outside the particles).
	Non-spherical particles allow the evaluation only outside the particles.
	The computed near fields inside the circumscribing sphere of non-spherical particles are in general not correct.
	
	\paragraph{Far fields}
	A far field is an intensity distribution $I(\alpha,\beta)$ in direction space (i.e., power per solid angle, measured far away from the scattering centers), such that the total power is:
	\begin{align}
	\label{eq:farfield}
	W=\int_0^{2\pi}\int_0^\pi I(\alpha,\beta)\sin\beta \dd{\beta} \dd{\alpha}
	\end{align}	
	The following code snippet illustrates several convenience functions to visualize the far field:
	
\begin{lstlisting}
from smuthi.postprocessing.graphical_output \
     import show_scattered_far_field, \
            show_total_far_field

show_scattered_far_field(simulation=simul)

show_total_far_field(simulation=simul)
\end{lstlisting}	
	Again, for a full list of possible settings, see the online documentation.

	\paragraph{Cross sections} If the initial field is a plane wave, the
	total or differential scattering cross section as well as the extinction
	cross section can be evaluated (see section \ref{sec:cross sections}).

\begin{lstlisting}
from smuthi.postprocessing.graphical_output \
     import show_scattering_cross_section
from smuthi.postprocessing.far_field \
     import total_scattering_cross_section, \
            extinction_cross_section

sc = total_scattering_cross_section(simulation=simul)

ec = extinction_cross_section(simulation=simul)

# differential cross section
show_scattering_cross_section(simulation=simul)
\end{lstlisting}

	\paragraph{Purcell factor}
	For simulations with a dipole source, the total radiated power can be evaluated with the following commands:	
\begin{lstlisting}
# after the simulation has been run
p0 = dip.dissipated_power_homogeneous_background(
           layer_system=simul.layer_system)

p = dip.dissipated_power(
           particle_list=simul.particle_list,
           layer_system=simul.layer_system)

purcell = p / p0
\end{lstlisting}
	The Purcell factor \cite{purcell1995spontaneous} is then defined as the power radiated into the system by the dipole, divided by the power that it would radiate in a homogeneous medium.
	
	If the user needs post-processing that goes beyond the described functionality, we recommend to browse through the online API documentation of the \lstinline!postprocessing! package or directly through the source code and construct your own post-processing routine from the provided data structure.
	
	\subsection{Numerical parameters}
	\label{sec:numpar}
	In addition to the parameters that define the optical system, several numerical parameters can be used to control the accuracy (and runtime) of a simulation.
	
	\subsubsection{Multipole cut-off}	
	\label{sec:cutoff}
	
	The scattering properties of each particle are represented by its T-matrix $T_{\tau lm,\tau'l'm'}$,
	where $\tau lm$ and $\tau'l'm'$ are the multipole polarization, degree and order of the scattered and incoming field, respectively.
	In practice, the T-matrix is truncated at some multipole degree $\lmax\geq 1$ and order $0\leq\mmax\leq\lmax$ to obtain a finite system of linear equations.
	
	The user can specify the cut-off parameters for each particle like this:

\begin{lstlisting}
large_sphere = Sphere( ...
                      l_max=@10@,
                      m_max=@10@,
                       ...)

small_sphere = Sphere( ...
                      l_max=@3@,
                      m_max=@3@,
                       ...)	
\end{lstlisting}
	In general, we can say:
	\begin{itemize}
		\item Large particles require higher multipole orders than small particles.
		\item Particles very close to each other, very close to an interface or very close to a point dipole source require higher multipole orders than those that stand freely.
		\item Larger multipole cutoff parameters imply better accuracy, but also a quickly growing numerical effort.
		\item Setting a too large multipole cutoff order can lead to numerical instabilities.
		\item When simulating flat particles near planar interfaces, the multipole truncation should be chosen with regard to the Sommerfeld integral truncation, compare \cite{egel2017extending}.
	\end{itemize}
	
	Several rules of thumb can be found in the literature for the selection of the multipole truncation in the case of spherical particles, see for example \cite{neves2012effect, wiscombe1980improved}.
	Their applicability to the case of particles near planar interfaces, however, is typically not guaranteed.
	For this reason, \smuthi offers a built-in automatic parameter selection feature to estimate a suitable multipole truncation, see \ref{sec:autoparam}.

	\subsubsection{Integral contours}
	
	Several steps during a \smuthi simulation involve the numerical evaluation of integrals over the in-plane wavenumber of a plane wave expansion	
	\begin{align}
	\label{eq:int}
	\int_0^\infty \hspace{-0.2cm} f(\kappa) \dd{\kappa},
	\end{align}
	where $\smash{\kappa=\sqrt{k_x^2+k_y^2}}$.
	To manage these integrals, we often refer to the dimensionless \emph{effective refractive index}
	\begin{align}
	\neff = \frac{\kappa\lambda}{2\pi},
	\end{align}
	where $\lambda$ is the vacuum wavelength.
	
	In order to avoid numerical instability due to waveguide-mode or branchpoint singularities, it is favorable to replace the integral \eqref{eq:int} along the real axis by a complex path $C:[0,1]\rightarrow\mathbb{C}$ that is slightly deflected into the negative imaginary (see for example section 2.7.3 of \cite{chew1995waves}):
	\begin{align}
	\label{eq:trunc}
	\int_0^\infty \hspace{-0.2cm} f(\kappa) \dd{\kappa} &\longrightarrow \int_C f(\kappa) \dd{\kappa},
	\end{align}
	
\begin{figure}
		\centering
		\includegraphics{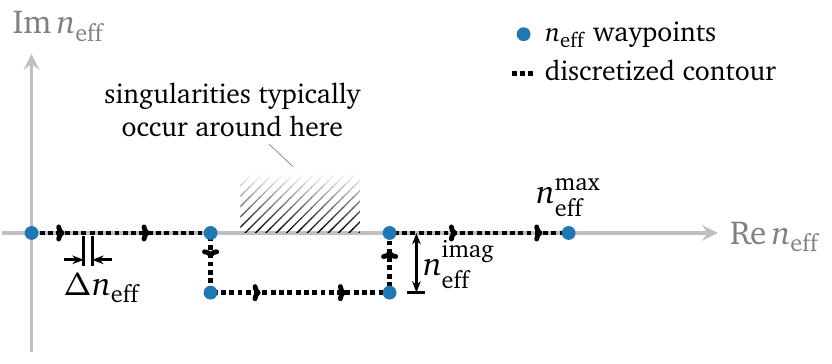}
		\caption{Parameters to control a deflected contour}
		\label{fig:contour}
	\end{figure}

	The following parameters can be used to define a deflected contour (compare figure \ref{fig:contour}):
	\begin{itemize}
		\item The truncation point $\neffmax$.
			  In a layer with refractive index $n$, all partial plane waves with $\neff<n$ are propagating and all waves with $\neff>n$ are evanescent.
			  The parameter $\neffmax$ therefore defines how deep into the evanescent region the field expansion should be considered.
			  As a rule of thumb, $\neffmax$ should be chosen well above the largest occurring refractive index, e.g., $\neffmax = n + 1.2$, which is the default.
			  Note that a dipole source or a small particle very close to a layer interface might require a larger $\neffmax$.
		\item The deflection into the imaginary $\neffimag$.
			  If small, a fine discretization $\dneff$ might be required because the integrand can then show rapid variations near waveguide mode or branchpoint singularities.
			  However, a too large deflection into the imaginary can cause numerical problems, especially if particles are separated by a very large lateral distance.
			  In that case, it is recommended to keep $\neffimag < 2/(k\rho_\mathrm{max})$, where $k=2\pi/\lambda$ is the vacuum wavenumber and $\rho_\mathrm{max}$ is the largest lateral separation between two particles, see section \ref{sec:distant}.
		\item The discretization along the path, $\dneff$.
			  If a small value is chosen, the integral is more accurate, but the runtime grows.
			  Note that for large lateral inter-particle distances, $\dneff$ must be chosen small in order to avoid aliasing.
			  Again, it is recommended to keep $\dneff < 2/(k\rho_\mathrm{max})$, see section \ref{sec:distant}.
	\end{itemize}

	\paragraph{Default contours}
	The simplest and recommended way to manage integral contours is through the globally defined \emph{default contours}.
	Two default contours exist: a default Sommerfeld integral contour (which is by default used to compute the layer system mediated scattered field) and a default initial field integral contour (which is by default used to evaluate the initial field in case of dipole or Gaussian beam excitation).
	These contours are automatically set in the beginning of a simulation run, unless the \lstinline!overwrite_default_contours! input argument of the simulation class is set to \lstinline!False!.
	That way, the input arguments of the \lstinline!Simulation! constructor can be used to set the contour parameters.
	Specifying, for instance:
\begin{lstlisting}
from smuthi.simulation import Simulation

simul = Simulation( ...
                   k_parallel="default",
                   neff_max=@2.5@,
                   neff_imag=@5e-3@,
                   neff_resolution=@2e-3@,
                   ... )
\end{lstlisting}
	forces the setting of default contours using the specified parameters at the beginning of the simulation run.
	If you don't specify these arguments, the default settings of the \lstinline!Simulation! class are applied.
	They are chosen such that for many application scenarios a reasonable integral contour is created.

	\begin{figure}[tb]
	\centering
	\includegraphics{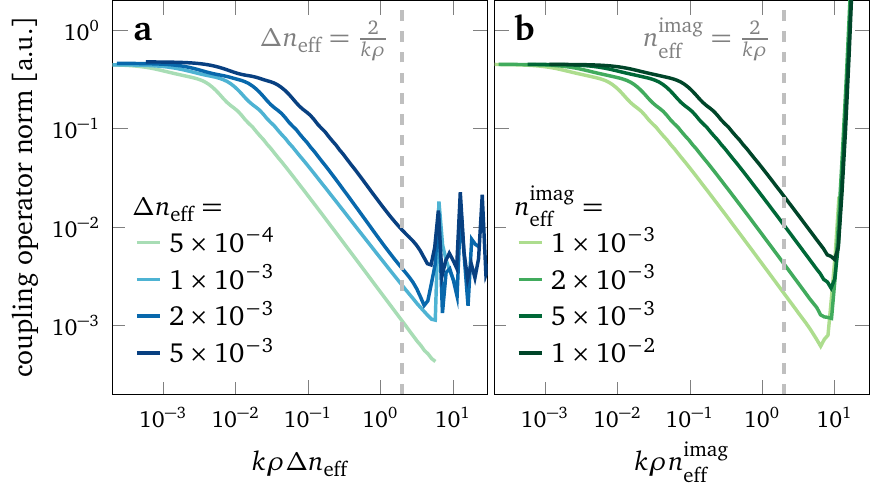}
	\caption{Aliasing and related issues for distant particles.
			 a) Calculated layer system mediated particle coupling operator as a function of dimensionless lateral particle separation $k\rho$ scaled by $\dneff$ for various values of $\dneff$.
			 b) The same for $\neffimag$ instead of $\dneff$.
			 See text for details.}
	\label{fig:aliasing}
	\end{figure}

\begin{figure}[htb]
	\centering
	\includegraphics{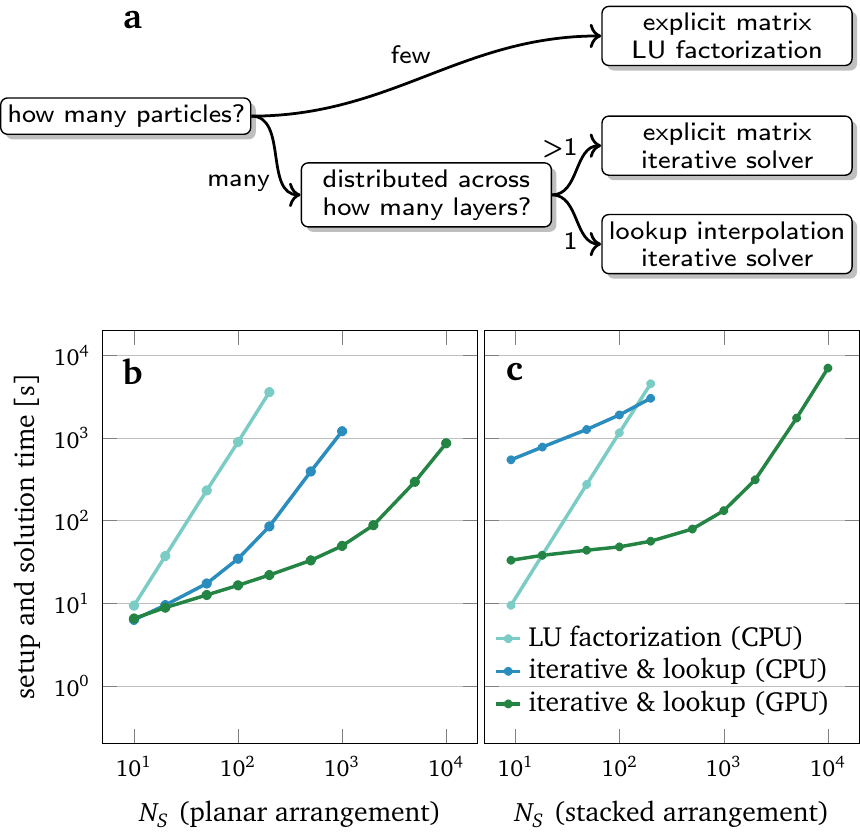}
	\caption{a) Suggested solver strategy among those available in \smuthi. b-c) Linear system setup and solution time as a function of particle number for identical glass spheres on a glass substrate. We compare direct inversion on CPU with iterative solution on CPU and on GPU. The data for iterative solutions were generated in combination with lookup interpolation. Creating and querying the lookup table is faster if all particles share the same $z$ coordinate (planar arrangement, panel b) as compared to arrangements where the particles are distributed over a span of heights (stacked arrangement, panel c). The actual runtime duration depends on the the numerical and solver settings, as well as on the computer hardware. Plotted data refer to simulations with $\lmax=\mmax=3$, run on a Google Colab \cite{colab} runtime equipped with an Intel Xeon CPU@\SI{2.30}{\giga\hertz} and a Tesla T4-16GB GPU.}
	\label{fig:solvers}
\end{figure}

\paragraph{Overriding the default contour}
Every object in a \smuthi simulation that defines integrals of type \eqref{eq:int} has a \sloppy \lstinline!k_parallel! or \lstinline!k_parallel_array! argument in its constructor.
If an array of $\kappa$ values is specified, that array is used instead of the default contour.

\begin{lstlisting}
from smuthi.fields \
     import reasonable_Sommerfeld_kpar_contour
from smuthi.initial_field import DipoleSource

# define suitable k_parallel array
kp_ar = reasonable_Sommerfeld_kpar_contour(
            vacuum_wavelength=@0.5@,
            layer_refractive_indices=[@1.52@,@1@],
            neff_imag=@1e-3@,
            neff_max=@5@,
            neff_resolution=@1e-3@)

# apply to specific dipole object
dip = DipoleSource( ...
                   k_parallel_array=kp_ar
                    ... )
\end{lstlisting}
	\fussy
	It is advisable to override the default contour if different objects have a different convergence behavior with regard to the contour parameters.
	For example, if a dipole source is placed very close to an interface, a very large $\neffmax$ might be required for convergence.
	However, when setting the default contours with such a large $\neffmax$ value, all Sommerfeld integrals (in particular the ones that govern the particle multiple scattering) will be evaluated with that setting.
	Then it is better to set a moderate $\neffmax$ for the default contour and override the contour in the dipole constructor separately.

	\subsubsection{Distant particles: Aliasing and related problems}	
	\label{sec:distant}
	
	Mathematically, Sommerfeld integrals are Hankel transforms, where the in-plane wavenumber $\kappa = 2\pi/\lambda \neff$ and the lateral separation $\rho = \sqrt{\Deltaup x^2 + \Deltaup y^2}$ constitute a transform pair.
	As a consequence, a large lateral separation implies the need for a fine resolution of the integrand in numerical quadrature -- otherwise aliasing errors are encountered.
	
	This issue is illustrated in figure \ref{fig:aliasing}a, where the layer mediated particle coupling strength $W^{i,R}_{j}$ for two particles above a glass substrate is shown as a function of dimensionless distance.
	Each curve refers to a different integrand sampling resolution $\dneff$, with the distance scaled by $\dneff$ to compare the different curves directly.
	For this figure, the deflection into the imaginary was fixed to $\neffimag=\num{2e-4}$.
	
	With growing distance, the coupling strength drops, because the electromagnetic interaction becomes weaker.
	However, for a certain large distance, the monotonous decay breaks and the onset of a noisy behavior marks the point where aliasing becomes relevant.
	We can identify $\dneff < 2/(k\rho)$ as a threshold, below which aliasing is not to be expected.
	
\begin{figure*}[tb]
	\centering
	\includegraphics{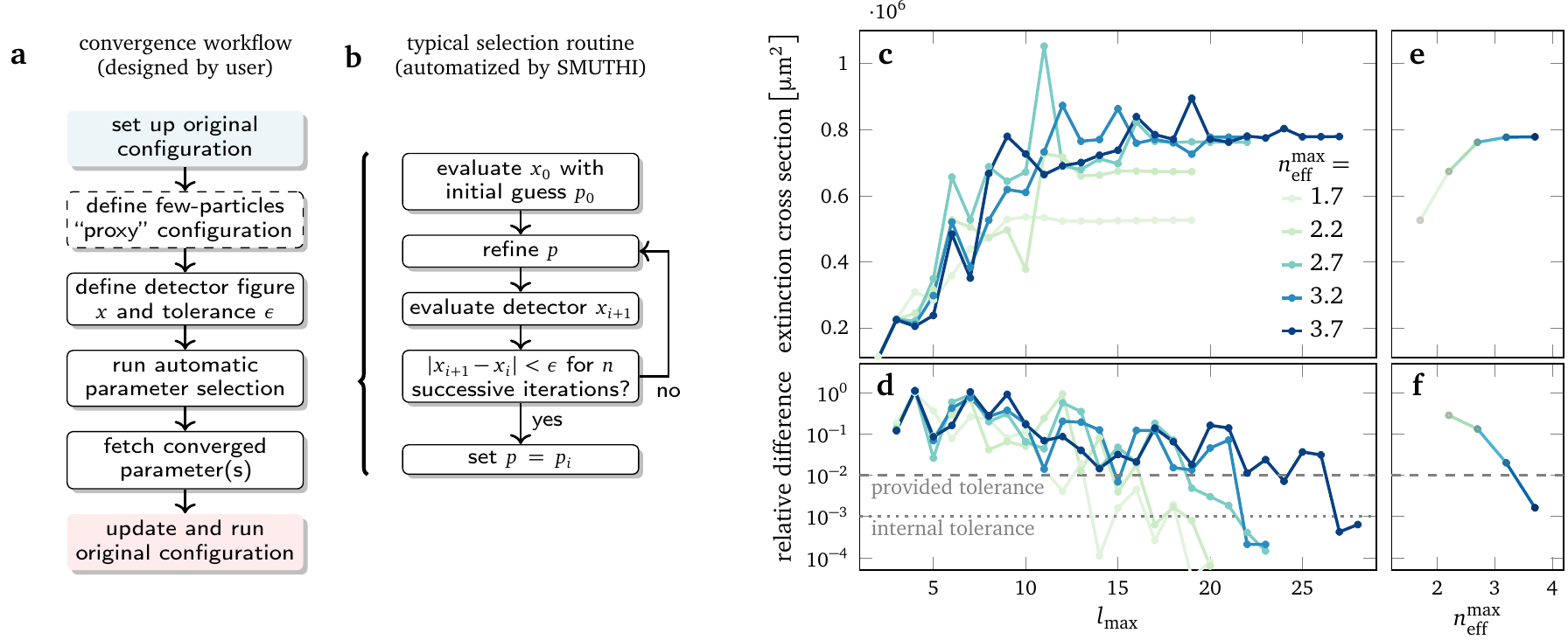}
	\caption{Overview of the automatic parameter selection process. a) Flowchart representation of the steps that the user takes in order to use the automatic parameter selection. The second step (definition of proxy-simulation) is optional, and only required if the original configuration involves a long run time, see section \ref{sec:proxy conf}. b) Flowchart of the typical algorithm followed to select a single numerical parameter. For further details, see the online documentation. c-f) Illustrative data generated during the selection of the appropriate $\neffmax$ value for a particle on a substrate using the extinction cross section as the detector figure $x$ and a tolerance of $\epsilon = \num{e-2}$.}
	\label{fig:autoparam}
\end{figure*}
	
	A similar issue affects the choice of the contour deflection into the imaginary, $\neffimag$, see \ref{fig:aliasing}b.
	Here, $\dneff$ is fixed to \num{e-4}.
	The reason for the numerical errors at large distances is that the quadrature of the Sommerfeld integrals becomes numerically unstable if the argument $k\rho\neff$ of the involved Bessel functions has a too large negative imaginary part.
	Again, a threshold of $\neffimag < 2/(k\rho)$ can be used to prevent aliasing.

	\subsubsection{Angular resolution}	
	When the far field quantities are evaluated in direction space, the angular resolution determines the accuracy of integrated power quantities, like the total power flux or the total scattering cross section, compare \eqref{eq:farfield}.
	Depending on the size and configuration of scattering particles, the scattered intensity as a function of direction can have narrow features (e.g., a sharp peak in forward direction for large particles).
	Then, a fine angular resolution might be necessary to obtain correct integral results.
	
	The default angular resolution can be set as an input argument of the \lstinline!Simulation! object:
	\begin{lstlisting}
from smuthi.simulation import Simulation
import numpy as np

simul = Simulation( ...
                    angular_resolution=np.pi / @360@
                    ...)
	\end{lstlisting}
	Note that the angular resolution only impacts the post processing, but has no influence on the steps performed during the \lstinline!Simulation.run()! command.	
	
	\subsubsection{Solver settings}
	In order to allow runtime-efficient simulations, \smuthi currently offers two numerical strategies for the solution of the scattering problem:
	\sloppy
	\begin{enumerate}
		\item LU factorization: To this end, the interaction matrix $M^i_{j,nn'}$ (see \eqref{eq:master}) is fully stored in memory.
		\item Iterative solution with the GMRES, LGMRES or GCROTMK method.
			  In this case, you can either store the full interaction matrix in memory, or use a lookup from which the matrix entries are approximated by interpolation, see section 3.10.1 of \cite{egel2019accurate} or \cite{egel2016efficient}.
	\end{enumerate}
	\fussy

	With growing particle number, all involved operations get more expensive, but the cost of LU factorization grows faster than the cost of iterative solution.
	Similarly, the calculation cost of the full interaction matrix grows faster than the cost of computing a lookup table.
	For this reason, we recommend the decision scheme depicted in figure \ref{fig:solvers}a. Figure \ref{fig:solvers}b and c illustrates the typical scaling of solver time as a function of particle number for different solver settings and hardware setups. 
	It is worth noting that the lookup interpolation approach is even more efficient for configurations where all particles are located at the same height, i.e., they share the same $z$-coordinate (figure \ref{fig:solvers}b) as compared to the general case where particles are distributed over a span of heights (\ref{fig:solvers}c).
	The plotted data also illustrates that for small particle numbers, the preparation of the lookup tables limits the runtime, whereas for large particle numbers, the iterative solution becomes the computational bottleneck. The cross-over between these regimes is marked by a change in the double-logarithmic slope of the corresponding graphs.
	
	In a \smuthi script, the numerical strategy for solving the linear system is defined through the input parameters of the simulation constructor.
	The relevant parameters are:
\sloppy	
	\begin{enumerate}
		\item \lstinline!solver_type!: At the moment, \lstinline!"LU"! (default), \lstinline!"GMRES"!, \lstinline!"LGMRES"! or \lstinline!"GCROTMK"! are available.
		\item \lstinline!solver_tolerance!: Convergence criterion (for iterative solver only, default \num{e-4}).
		\item \lstinline!store_coupling_matrix!: If true (default), the coupling matrix is explicitly calculated and stored in memory.
			  Otherwise, a lookup table is prepared and the matrix-vector multiplications are run on the fly, where the matrix entries are computed using the lookup table.
			  Note that lookup table interpolation is currently only available if all particles are in the same layer.
			  The parameter is ignored in case of LU solver type.
		\item \lstinline!coupling_matrix_lookup_resolution!: In case lookup tables are used, this sets the sampling step distance (in length units).
			  The parameter is ignored when the coupling matrix is explicitly calculated.
		\item \lstinline!coupling_matrix_interpolator_kind!: In case lookup tables are used, switch between \lstinline!"linear"! or \lstinline!"cubic"! (default) interpolation.
			  \lstinline!"linear"! is faster and \lstinline!"cubic"! is more precise on equal resolution.
			  The parameter is ignored when the coupling matrix is explicitly calculated.
	\end{enumerate}
\fussy
	
	This would be a typical setting for a small number of particles:
\begin{lstlisting}	
from smuthi.simulation import Simulation

simul = Simulation( ...
                   solver_type="LU",
                   store_coupling_matrix=@True@,
                    ... )
\end{lstlisting}	
	This would be a typical setting for a large number of particles:
\begin{lstlisting}	
from smuthi.simulation import Simulation

simul = Simulation( ...
         solver_type="GMRES",
         solver_tolerance=@1e-3@,
         store_coupling_matrix=@False@,
         coupling_matrix_lookup_resolution=@5@,
         coupling_matrix_interpolator_kind="linear",
                    ... )
\end{lstlisting}	
	Note that GPU acceleration is currently only available for particle coupling through lookup interpolation.

	\subsection{Automatic parameter selection}
	\label{sec:autoparam}
	
	Smuthi offers a module to run an automated convergence test for the following parameters:
	\begin{itemize}
		\item Multipole truncation parameters $\lmax$ and $\mmax$ for each particle
		\item Integral contour parameters $\neffmax$ and $\dneff$
		\item The angular resolution $\Deltaup\beta$, $\Deltaup\alpha$ of far field data
	\end{itemize}	
	The user provides: a simulation object, a detector function and a relative tolerance.
	The detector function maps a simulation object (that has already been run) to one numeric figure (the detector quantity).
	In other words, the detector function does some post-processing to yield a scalar value that we use to monitor convergence.
	The following pre-defined strings can be passed as arguments for convenience: \lstinline!"extinction cross section"!, \lstinline!"total scattering cross section"! or \lstinline!"integrated scat- tered far field"! to set the corresponding figure as the detector quantity.
	Other possible detector functions could map to the norm of the electric field at a certain point, or the scattered far field in a certain direction or whatever appears be a suitable measure for the convergence of the simulation.

	The automatic parameter selection routine then repeatedly runs the simulation and evaluates the detector quantity with subsequently modified numerical input parameters until the relative deviation of the detector quantity is less than the specified tolerance, see figure \ref{fig:autoparam}. 

	After termination, the simulation object is updated with the so determined convergence settings. The following code example illustrates a possible use of the feature:
\begin{lstlisting}
from smuthi.utility.automatic_parameter_selection
     import select_numerical_parameters

select_numerical_parameters(simulation=simul,
                            tolerance=@1e-3@)
\end{lstlisting}	
		
	Some aspects need to be taken into account when using the automatic parameter selection:
	\begin{itemize}
		\item Both the multiple scattering and the initial field default contour are updated with the same parameters.
			  A separate optimization of the parameters for initial field and multiple scattering is currently not supported.
		\item The algorithm compares the detector value for subsequent simulation runs.
			  The idea is that if the simulation results agree for different numerical input parameters, they have probably converged with regard to that parameter.
			  However, in certain cases this assumption can be false, i.e., the simulation results agree although they have not converged.
			  The automatic parameter selection therefore does not replace critical judging of the results by the user.
		\item With the parameter \lstinline!tolerance_steps!, the user can ask that the tolerance criterion is met multiple times in a row before the routine terminates.
		\item The simulation is repeated multiple times, such that the automatic parameter selection takes much more time than a single simulation.
		\item For flat particles on a substrate, it is recommended to provide \lstinline|relative_convergence=True| as an input argument to the \lstinline|select_numerical_parameters| method. This triggers a separate convergence run for the multipole cutoff parameters during each iteration of the $\neffmax$ parameter to account for the cross-dependent convergence behavior of these two sets of parameters, see \cite{egel2016light, egel2017extending}.		
	\end{itemize}
	For further details, see the online documentation.
	
	\subsubsection{Automatic parameter selection for simulations with many particles}
	\label{sec:proxy conf}
	A simulation with many particles can be busy for a considerable runtime.
	The above described automatic procedure might then be unpractical.
	In this case, we recommend to define a ``proxy configuration''.
	The idea is to find a system that takes less time to simulate but that has similar requirements with regard to numerical parameters.

	Let us for example assume that we want to simulate light scattering by one thousand identical flat nano-cylinders located on a thin film system covering a substrate.
	Then, the selection of $\neffmax$ needs to be done with regard to the distance of the particles to the next planar interface, whereas $\lmax$ and $\mmax$ have to be chosen with regard to the particle geometry, material, and to the selected $\neffmax$.
	Finally, $\dneff$ needs to be chosen with regard to the layer system response.
	All of these characteristics have nothing to do with the fact that we are interested in a many particles system\footnote{One does, however, have to consider the implications of large lateral inter-particle distances on the choice of $\dneff$ and $\neffimag$, see section \ref{sec:distant}}.
	We can thus simulate scattering by a single cylinder (or, to be on the safe side, by two neighboring cylinders) on the thin film system and let the automatic parameter selection module determine suitable values for $\lmax$, $\mmax$, $\neffmax$ and $\dneff$.
	These parameters are then used as input parameters for the \num{1000}-particles simulation which we run without another call to the automatic parameter selection module.
	
	\subsection{Physical units}
	
	\smuthi is committed to a ``relative-units'' philosophy.
	That means, all quantities have only relative meaning.

	\paragraph{Length units}	
	The user must consistently provide all length parameters (particle sizes, layer thicknesses, wavelengths, lookup resolutions, etc.) using a length unit of choice.
	Results will refer to that same unit.
	For example, if you specify the wavelength in nanometers, resulting cross sections will be in \si{\nano\meter\squared}.
	Quantities with an inverse length dimension (wavenumbers) also implicitly refer to the selected length unit.

	\paragraph{Field strength units}	
	When the electromagnetic fields are computed, their absolute value has no physical meaning.
	Only relative quantities can be used for further analysis.
	For example, the scattered field strength divided by the amplitude of the initial field does have a physical meaning.

	\paragraph{Power units}	
	Also power units have no meaning as absolute values.
	To get meaningful information, power-related figures always need to be regarded with reference to other power-related figures.
	Some examples:
	\begin{itemize}
	\item Scattering cross section as the quotient of scattered (angular) intensity and incident (power per area) intensity.
	\item Diffuse reflectivity as the total back scattered far field power divided by the initial Gaussian beam power.
	\item Purcell factor as the dissipated power of a dipole source divided by the dissipated power of the same source in an infinite homogeneous medium (i.e., in the absence of planar interfaces and scattering particles).
	\end{itemize}
	
	\subsection{Cross sections}
	\label{sec:cross sections}
	If the initial excitation is given by a plane wave, it is natural to discuss the far field properties of a scattering structure in terms of cross sections.
	However, in the context of scattering particles near planar interfaces, the commonly used concepts of cross sections need further clarification.
	In the following, we therefore discuss the meaning of cross sections as they are implemented in \smuthi.

\begin{figure}
	\centering
	\includegraphics{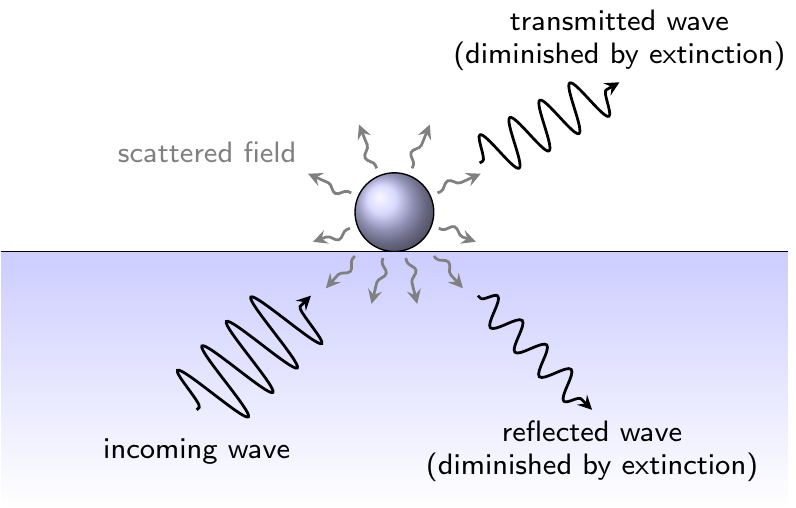}
	\caption{Concept of extinction in the presence of interfaces (compare \cite{egel2019accurate}).}
	\label{fig:extinction}
\end{figure}

	\subsubsection{Scattering cross section}	
	The concept of a scattering cross section is straightforward: The incoming wave is specified by an intensity (power per area), whereas the scattered field is characterized by a power, such that the scattered signal divided by the initial signal yields an area.
	
	The total scattering cross section reads
	\begin{align}
		C_\scat=W_\scat I_\init
	\end{align}	
	where $W_\scat$	is the total scattered power and $I_\init$ is the incident irradiance (power per unit area perpendicular to the direction of propagation).

\begin{figure}[tb]
	\centering
	\includegraphics{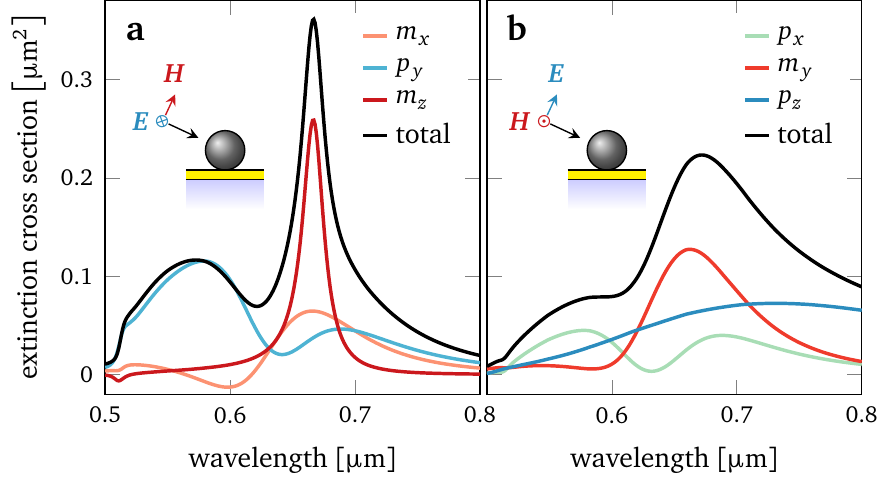}
	\caption{Multipole decomposition of the extinction cross section for a silicon sphere on a gold-coated glass substrate, illuminated by a) a TE-polarized and b) a TM-polarized wave propagating at an angle of \ang{65}. Compare figure 3(b,d) of \cite{sinev2016polarization}. In the legend, $p_x, p_y, p_z$ $(m_x, m_y, m_z)$ refer to the Cartesian components of the electric (magnetic) dipole contributions.}
	\label{fig:multipoles}
\end{figure}%

\begin{figure*}[tb]
	\centering
	\includegraphics{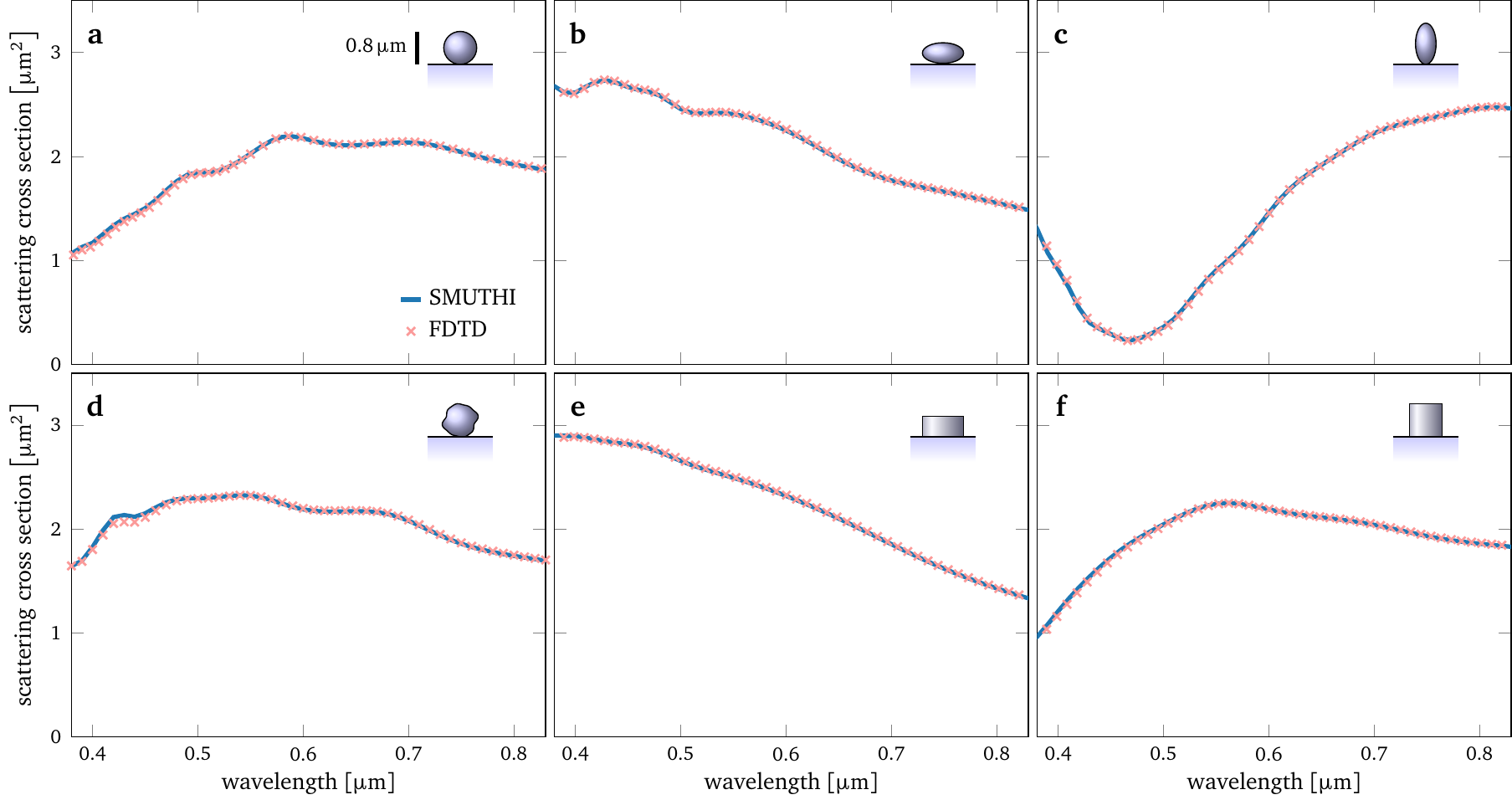}
	\caption{Scattering cross section spectra for a single glass particle on a glass substrate: a) sphere, b) prolate spheroid, c) oblate spheroid, d) free-form particle, e) flat cylinder, f) cylinder with unit aspect ratio.
		All particles have the same volume.}
	\label{fig:dielsphonsubstrate}
\end{figure*}%

	\begin{table*}[htb]
	\setlength{\tabcolsep}{5pt}
	\caption{Description of geometric and simulation parameters used to generate the data of figure \ref{fig:dielsphonsubstrate}. Particle parameters $p_i$ correspond to the radius for the \lstinline!Sphere!, the two semi-axes for the \lstinline!Spheroid!, and the radius and height for the \lstinline!FiniteCylinder!, respectively.}
	\label{tab:examples}
	\footnotesize
	\centering
	\resizebox{\linewidth}{!}{
	\begin{tabular}{c c c c c
	                *{3}{S[table-format=1.3]}
	                *{1}{S[table-format=1.2]}
	                *{2}{S[table-format=1.0]}
	                *{1}{S[table-format=1.1]}
	                *{1}{S[table-format=1.2]}
	                *{1}{S[table-format=2.0]}
	                *{1}{S[table-format=1.0]}
	                *{2}{S[table-format=1.2]}
	               }
	\toprule
	{\multirow{3}{*}{case}}
		& \multicolumn{3}{c}{initial field}
		& \multicolumn{5}{c}{particle parameters}
		& \multicolumn{4}{c}{layer system}
		& \multicolumn{4}{c}{simulation parameters} \\
	\cmidrule(lr){2-4}
	\cmidrule(lr){5-9}
	\cmidrule(lr){10-13}
	\cmidrule(lr){14-17}
	   & {type}	& {$\theta$} 	   & {pol} & {type} & {$p_1$} & {$p_2$} & {$z$} & {$n$} & {$t_1$} & {$t_2$} & {$n_1$} & {$n_2$} & {$\lmax$} & {$\mmax$} & {$\neffmax$} & {$\dneff$} \\
	   & & {[\si{\radian}]} & & & {[\si{\micro\meter}]} & {[\si{\micro\meter}]} & {[\si{\micro\meter}]} & & {[\si{\micro\meter}]} & {[\si{\micro\meter}]} & & & & & & \\
	\midrule
	 a & \lstinline!PlaneWave! & $\pi$ & TE & \lstinline!Sphere!			& 0.4   		& {\textemdash}	& 0.4   & 1.52 & $\infty$ & $\infty$ & 1.0 & 1.52 & 9  & 1 & 2.22 & 0.01 \\
	 b & \lstinline!PlaneWave! & $\pi$ & TE & \lstinline!Spheroid!			& 0.504 		& 0.252 		& 0.252 & 1.52 & $\infty$ & $\infty$ & 1.0 & 1.52 & 23 & 1 & 2.22 & 0.01 \\
	 c & \lstinline!PlaneWave! & $\pi$ & TE & \lstinline!Spheroid!			& 0.317 		& 0.635 		& 0.635 & 1.52 & $\infty$ & $\infty$ & 1.0 & 1.52 & 13 & 1 & 2.22 & 0.01 \\
	 d & \lstinline!PlaneWave! & $\pi$ & TE & \lstinline!CustomParticle!	& {\textemdash}	& {\textemdash}	& 0.326 & 1.52 & $\infty$ & $\infty$ & 1.0 & 1.52 & 10 & 3 & 2.42 & 0.01 \\
	 e & \lstinline!PlaneWave! & $\pi$ & TE & \lstinline!FiniteCylinder!	& 0.44			& 0.44			& 0.22	& 1.52 & $\infty$ & $\infty$ & 1.0 & 1.52 & 27 & 1 & 2.22 & 0.01 \\
	 f & \lstinline!PlaneWave! & $\pi$ & TE & \lstinline!FiniteCylinder!	& 0.35			& 0.7			& 0.35	& 1.52 & $\infty$ & $\infty$ & 1.0 & 1.52 & 19 & 1 & 2.22 & 0.01 \\
	\bottomrule
	\end{tabular}
		}
\end{table*}	

\begin{figure*}[tb]
	\centering
	\includegraphics{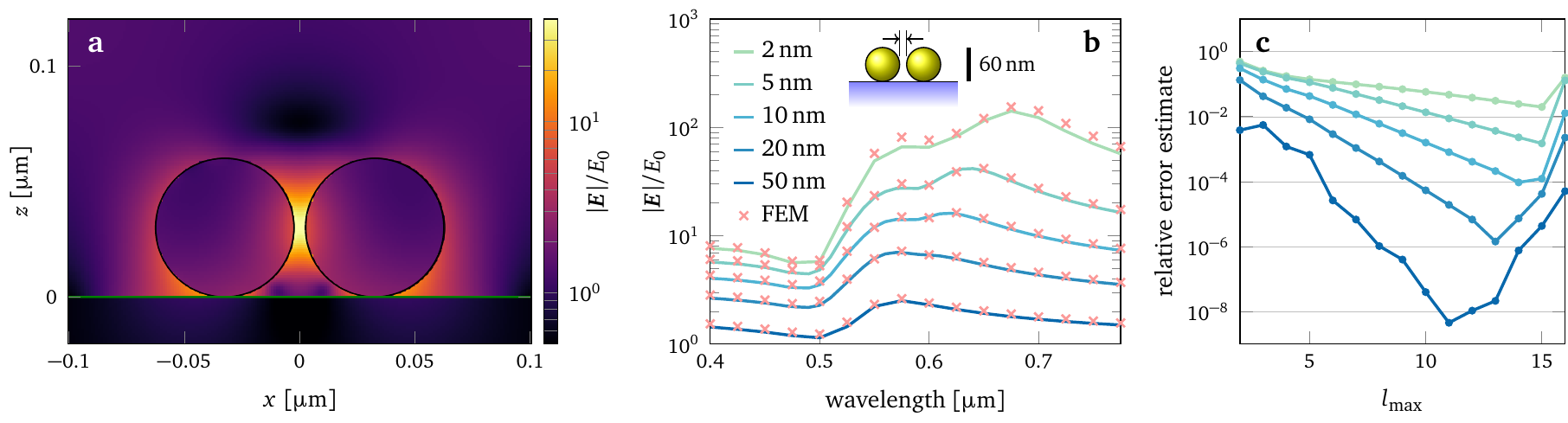}
	\caption{Gold dimer on a silicon substrate. a) Electric field distribution obtained for a plane wave illumination from above.
		The gap distance is \SI{5}{\nano\meter}.
		b) Electric field in the middle of the gap as a function of wavelength for various gap distances.
		c) Error estimate of the electric field in the gap as a function of multipole degree truncation for various gap distances.}
	\label{fig:sers}
\end{figure*}

	\subsubsection{Extinction cross section}	
	The term ``extinction'' means that particles take away power from the incident plane wave, such that they partially extinguish the incident wave.
	The power that they take away from the incoming wave is either absorbed or scattered into other channels, such that in the context of scattering of a plane wave by particles in a homogeneous medium, the extinction cross section is usually defined as the sum of the total scattering cross section and the absorption cross section.
	
	However, this interpretation of extinction (i.e., the sum of particle absorption plus scattering) is not straightforward when a planarly layered medium is also involved.
	The reason is that besides particle absorption and scattering, also absorption and waveguiding in the layered medium has to be taken into account.
	
	For this reason, we apply what is usually referred to as the optical theorem to define extinction (see section 3.8.1 of \cite{egel2019accurate} for the mathematical details), by interpreting the term ``extinction'' strictly as a measure for \emph{how much power is taken away by the particles from the incident plane wave}.
	
	In fact, \smuthi internally computes two extinction cross sections: one for the reflected incoming wave and one for the transmitted incoming wave, see figure \ref{fig:extinction}.
	That means, the extinction cross section for reflection (transmission) refers to the destructive interference of the scattered signal with the specular reflection (transmission) of the initial wave.
	It thereby includes absorption in the particles, scattering, and a modified absorption by the layer system, e.g., through coupling into guided modes.
	
	As a consequence, the extinction cross sections can be negative if (for example due to a modified absorption in the layer system) more light is reflected (or transmitted) in the specular direction than would be without the particles.
	
	Conservation of energy is then expressed by the following statement: ``For lossless particles near or inside a lossless planarly layered medium (that doesn't support any waveguide modes), the sum of top and bottom extinction cross section equals the total scattering cross section''.

	\subsubsection{Multipole decomposition}

	As \smuthi builds on the expansion of the scattered field in spherical vector wave functions, it naturally lends itself to a multipole analysis of the extinction cross section%
\footnote{A similar decomposition of the scattering cross section is in general not reasonable, because the scattered field intensity is not a linear function of the scattered field coefficients $b_n^i$, compare \eqref{eq:expansion1}. 
	For multiple particles or particles near planar interfaces, the contributions of individual multipole moments to the scattered field intensity is therefore not additive.}, see figure \ref{fig:multipoles}. 
\sloppy
	A convenience function to select the contribution of individual (spherical) multipole moments to the evaluation of the extinction cross section is built into the \lstinline|extinction_cross_section| method of the\linebreak \lstinline|smuthi.postprocessing.far_field| module. 
	See the online documentation for further details.
\fussy	

	\subsection{Restrictions} 
	The following restrictions limit the range of applications for \smuthi: 
	
	\begin{itemize}
		\item All media are linear, homogeneous, isotropic and non-magnetic. 
		\item Particles must not intersect with each other or with layer interfaces.
		\item The electric field inside the circumscribing sphere of a particle (i.e., the smallest sphere around a particle center that includes the particle volume) cannot be computed. 
		\item The software was designed for particles with diameters in the range of up to a few wavelengths. It's overall numerical stability and validity has not been tested for larger particles.
		\item Closely adjacent non-spherical particles with intersecting circumscribing spheres can lead to incorrect results.
		The use of \smuthi is therefore limited to geometries with particles that have disjoint circumscribing spheres.
		We note that by the use of regularized particle coupling approaches, this limitation can be relaxed \cite{theobald2017plane,martin2019t}.
		\item Dipole sources must not be placed inside the circumscribing sphere of a particle.
		\item If oblate particles are located near interfaces, such that the circumscribing sphere of the particle intersects the interface, a correct simulation result can in principle be achieved \cite{egel2016light}.
		However, special care has to be taken \cite{egel2017extending}.
		\item The software was designed for thin-film systems with layer thicknesses of up to a few wavelengths.
		Simulations involving thick layers might fail or return wrong results due to numerical instability.
	\end{itemize}
	
	In general, \smuthi does not provide error checking of user input, nor does it check if the numerical parameters specified by the user are sufficient for accurate simulation results or if the specified model falls into the scope of simulation scenes for which \smuthi can compute valid results.
	It is thus required that the user develops some understanding of the influence of various numerical parameters on the validity of the results and also for the limits of \smuthi's capabilities.	

	\section{Use case examples}
	\label{sec:examples}

	In the following, we present a number of representative example configurations that can be studied with \smuthi.
	For selected examples, we will also compare our results to accurate third-party benchmark simulations or results published in literature.
	Scripts to reproduce the results can be downloaded from the examples section in \smuthi's online git repository.

	\subsection{Single particle on glass substrate}
	
	A typical use case is the scattering of a plane wave by a single particle on a substrate.
	In this application example, we investigate the scattering and extinction cross section as a function of wavelength for dielectric particles ($n=\num{1.52}$) of different shapes on a dielectric substrate ($n=\num{1.52}$). See the caption of figure \ref{fig:dielsphonsubstrate} for further details.
	The particles are sized such that the equivalent volume radius is \SI{400}{\nano\meter}.
	Results are compared to FDTD data obtained using Lumerical \cite{lumerical}.
	
	Numerical parameters used for \smuthi simulations are determined by means of the automatic parameter selection feature, see section \ref{sec:autoparam}.
	To this end, we run the automatic parameter selection for the smallest wavelength of the considered spectral range and use the resulting numerical parameters for all wavelengths.
	The underlying assumption is that the smallest wavelength implies the largest particle size parameter, such that the corresponding numerical settings are a conservative choice for the other wavelengths, too.
	The automatic parameter selection is called with the default relative accuracy tolerance of \num{e-3}.
	Figure \ref{fig:dielsphonsubstrate} displays the spectral cross section for illumination by a plane wave under normal incidence.

	The automatic parameter selection resulted in the settings as summarized in table \ref{tab:examples}.
	The low value returned for $\mmax$ is typical for single axisymmetric particles illuminated under normal incidence.

	In all cases, the agreement to FDTD results is very good.

\begin{figure}[t]
	\centering
	\includegraphics{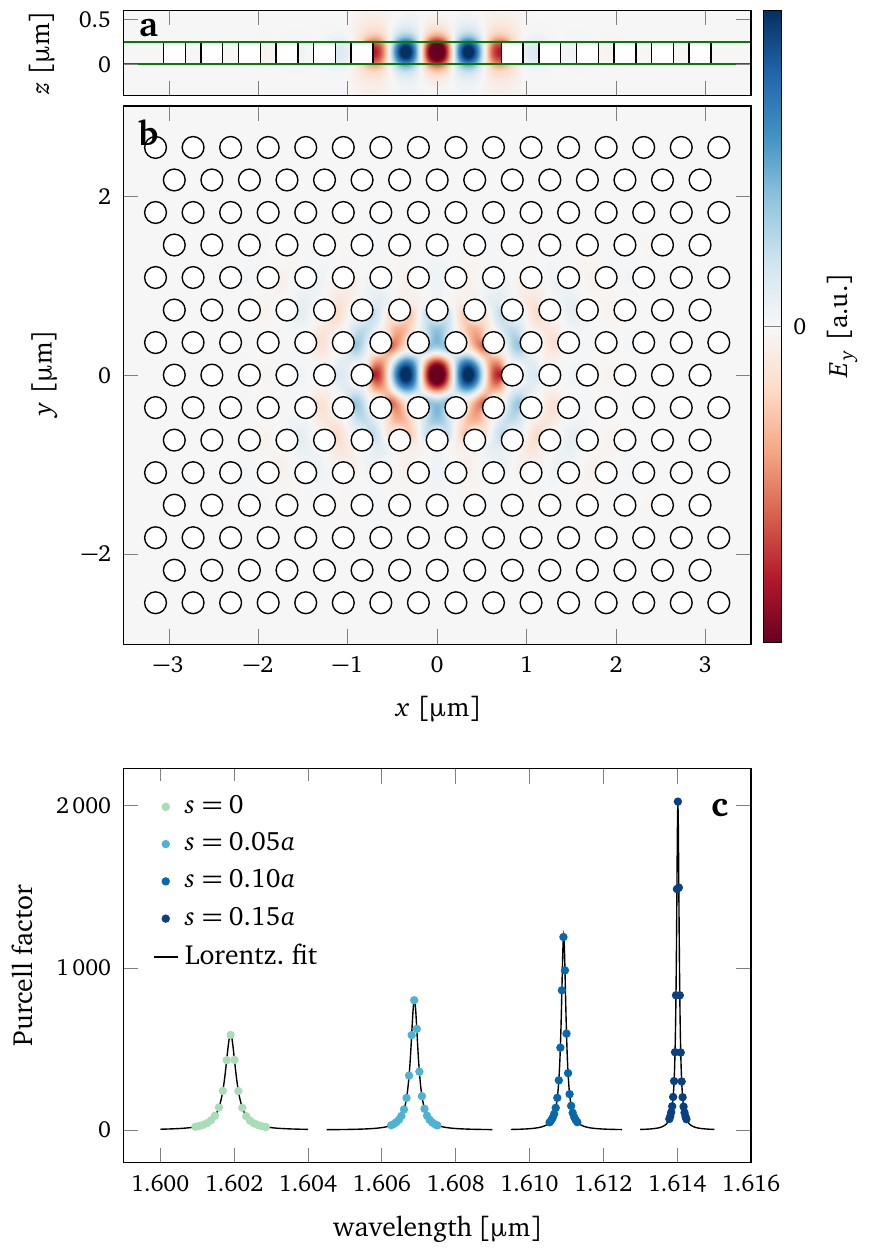}
	\caption{Field enhancement in a photonic crystal cavity. a-b) Electric field distribution for a $y$-aligned dipole source in a suspended photonic crystal membrane at resonance wavelength \SI{1.602}{\micro\meter}). Plotting planes correspond to $y=\SI{0}{\micro\meter}$ and $z=t/2$ where $t$ is the slab thickness.
		c) Purcell factor as a function of wavelength, compare figure 4 of \cite{akahane2003high}.}
	\label{fig:pcs}
\end{figure}

	\subsection{Field enhancement between two plasmonic nano spheres on a substrate}

	Electric field hot spots are important for non-linear applications such as Raman spectroscopy \cite{ding2016nanostructure}.
	In this example we want to explore how \smuthi can be used to analyze plasmonic structures with field enhancement.
	Two gold nanospheres (radius \SI{30}{\nano\meter}) are placed in water ($n=\num{1.33}$) on a silicon substrate.
	A gap of width $d$ defines the distance between the particles, compare figure 3c of \cite{ding2016nanostructure}.
	The scene is illuminated by a plane wave under normal incidence from above, with a polarization such that the electric field is parallel to the distance between the particle centers.
	Figure \ref{fig:sers}a shows the norm of the resulting electric field with a logarithmic color map for a gap width of $d=\SI{5}{\nano\meter}$.
	Panel \ref{fig:sers}b shows the norm of the electric field in the middle of the gap compared against results obtained via a finite element method using COMSOL \cite{comsol}.

	Figure \ref{fig:sers}c shows the relative difference of the calculated gap fields for subsequent values of the multipole degree truncation $\lmax$.
	We interpret this difference as a measure for the relative error of the gap field at that multipole truncation.
	As expected, the error decreases with growing $\lmax$.
	Configurations with a very narrow gap converge more slowly than configurations with a wider gap.
	At some threshold value, a numerical instability prevents further convergence, and the results become instead less accurate with growing $\lmax$.
	As a consequence, a relative accuracy of \SI{1}{\percent} is achieved at $\lmax=2$ for $d=\SI{50}{\nano\meter}$, at $\lmax=5$ for $d=\SI{20}{\nano\meter}$, at $\lmax=8$ for $d=\SI{10}{\nano\meter}$, at $\lmax=11$ for $d=\SI{5}{\nano\meter}$ and never for gap widths $d \leq \SI{2}{\nano\meter}$.

\begin{figure*}[t]
	\centering
	\includegraphics{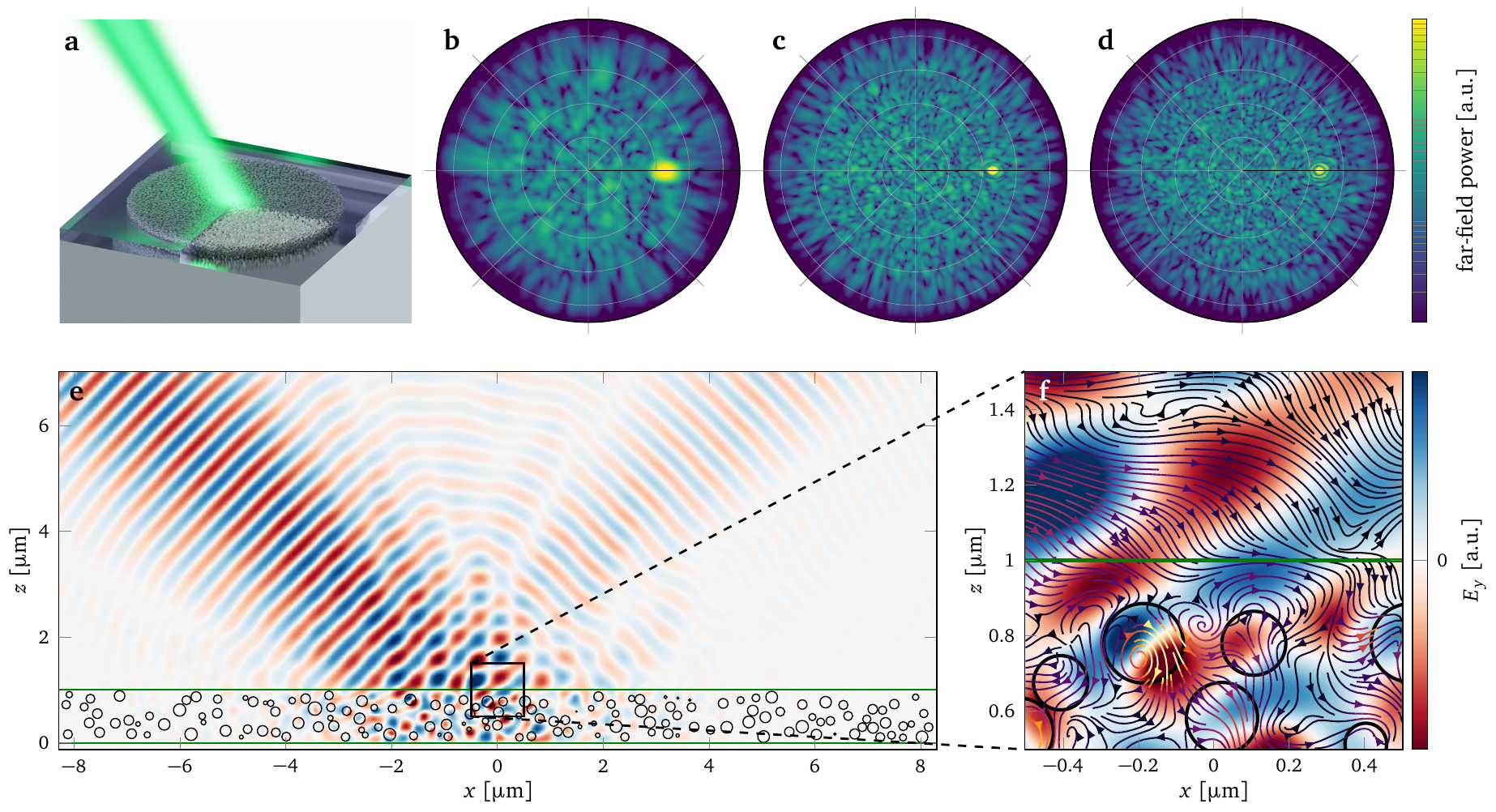}
	\caption{Tilted Gaussian beam impinging on a paint micro-layer over an iron substrate. a) Rendering of the tested configuration comprising \num{e4} particles. b-d) Reflected far-field power for incident beam waist values of \SIlist{4;6;8}{\micro\meter}, respectively. The specular reflection is visible as a bright spot in the intensity distribution. e) $E_y$ component of the electric field at the $y=\SI{0}{\micro\meter}$ plane. f) Detail of the near field distribution at the air-paint layer interface, with superimposed Poynting vector flux lines.}
	\label{fig:beam reflection}
\end{figure*}

	\subsection{Photonic crystal slab}
	In this application example, we reproduce the experimental findings presented in \cite{akahane2003high}.
	The geometry is given by a \SI{0.25}{\micro\meter} thick silicon slab with cylindrical air holes of radius \SI{0.12}{\micro\meter}.
	The air holes are arranged in a hexagonal grid with lattice constant $a=\SI{0.42}{\micro\meter}$.
	A defect of three missing air holes in a row acts as a photonic nano-cavity (known in the literature as the ``L3 cavity'').
	By shifting the air holes next to the defect location by a distance $s$ into the outward direction, the quality factor $Q$ of the cavity is optimized with a maximum for $s=0.15\,a$ \cite{akahane2003high}.
	
	We have reproduced the configuration in \smuthi, assuming a constant silicon refractive index of $n=\num{3.473}$ \cite{li1980refractive} and modeling the cavity by a rectangular domain of air holes with a wall ``thickness'' of seven lattice units in each direction, see figure \ref{fig:pcs}a-b, for a total of \num{230} air holes.
	The initial field is given by a point dipole source placed at the cavity center with a dipole moment aligned in the $y$-direction.
	In order to find suitable numerical settings, we used a proxy configuration (see section \ref{sec:proxy conf}) including only two air holes and launched an automatic parameter selection.
	The resulting parameters ($\lmax=5$, $\mmax=2$, $\neffmax=\num{5.67}$) were then used for the full simulations of the photonic crystal slab.
	
	For wavelengths close to a high-$Q$ resonance (in particular, wavelengths close to the peak resonance of \SI{1.1614}{\micro\meter} in the $s=0.15\,a$ case), the iterative solvers exhibited a slower convergence to the accurate solution -- sometimes stagnating altogether or converging to unphysical results.
	In the most critical case, convergence was eventually achieved using the \lstinline|GCROTMK| solver \cite{de1999truncation} as implemented in the \lstinline|scipy.sparse.linalg| library and relaxing the relative tolerance to \num{3e3}.
	Despite the difficulties to achieve convergence with low tolerance values, the resulting Purcell factors can be perfectly fitted by a Lorentzian lineshape, as expected.

	The upper panels of figure \ref{fig:pcs} show different crosscuts of the simulated electric field distribution, which show excellent agreement with numerical results published in the literature for the L3 cavity \cite{akahane2003high}.
	The Purcell factor as a function of wavelength is reported in the lower panel.
	When going from shift parameter $s=0$ to $s=\num{0.15}\,a$ in steps of $\num{0.05}\,a$, the subsequentially narrower resonances with higher amplitudes indicate a larger $Q$-factor of the cavity.
	At the same time, the resonance wavelength is shifted to longer wavelengths.
	These trends are again in good qualitative agreement with experimental data (compare figure 4 of \cite{akahane2003high}).

	\subsection{Beam reflection by paint micro layer on metal surface}
	A final example refers to a Gaussian beam that is incident on an iron surface covered with a scattering layer under an angle of \SI{45}{\degree}.
	The scattering layer has a thickness of \SI{1}{\micro\meter} and a refractive index of \num{1.52}.
	A total of \num{e4} polydisperse spherical TiO\textsubscript{2} particles are randomly dispersed in the scattering layer with a volume density of \SI{20}{\percent} over an area of \SI{\sim200}{\micro\meter\squared} (see figure \ref{fig:beam reflection}a).
	The sphere radii follow a Gaussian distribution around $r=\SI{0.1}{\micro\meter}$ with a standard deviation of $\sigma_r=\SI{0.01}{\micro\meter}$.
	Panels b-d show the far field intensity distribution obtained for increasing beam waist values, exhibiting an intense reflection peak in the specular direction. Panels e-f show the electric field $E_y$ at a cross-cut plane through the scattering layer, with a detail of the Poynting vector flux lines across the layer interface.

	\subsection{Further examples published in the literature}
	In addition to the presented use case examples, the interested reader can refer to the literature for additional applications where \smuthi has been used during its developing stages. Published works cover various systems in physics and electrical engineering research, including the design of nanoparticle based spectrally selective reflector layers in color conversion films \cite{theobald2020design},
	scattering layers for light outcoupling from organic light emitting diodes \cite{egel2017accurate},
	resonance analysis of spherical particles on a coated substrate \cite{pidgayko2020polarization},
	photon excitation from electron inelastic tunneling near a metallic or dielectric nano-sphere \cite{dvoretckaia2020electrically},
	the study of disordered meta surfaces \cite{czajkowski2019electromagnetic, czajkowski2020multipole}
	and the design of plasmonic gratings for sensing applications \cite{warren2020design}.

	\section{Conclusions and outlook}
	\label{sec:conclusions}
	
	In this paper, we have introduced a Python package for the simulation of electromagnetic scattering by multiple objects near or inside a planarly layered medium.
	After a brief overview on the theoretical background, we have provided a short manual highlighting the code interface from a user perspective.
	
	The accuracy of the software was validated by comparison to FDTD results for the illustrative use case of extinction by various particle shapes on a dielectric substrate.
	
	Near field enhancement for a plasmonic system of two metal spheres at variable distance on a silicon substrate was also evaluated, showing good agreement against FEM results down to a gap size of \SI{5}{\nano\meter} at optical wavelengths.
	We conclude that although the evaluation of the near field is limited to the domain outside the circumscribing sphere of a particle, \smuthi can be potentially used for near field studies of selected plasmonic structures.
	
	Further, we have studied the resonant behavior of a high-Q cavity in a silicon photonic crystal slab.
	To our knowledge, the simulation of photonic crystal slabs has not previously been demonstrated with the T-matrix method (related approaches that rely on the expansion of the scattered field in cylindrical waves rather than spherical waves have been published in \cite{boscolo2004three,pissoort2007fast}).	
	Due to its ability to handle large numbers of scatterers in contact with layer interfaces, we believe that \smuthi is a particularly powerful tool for this class of applications, which includes relevant platforms such as perforated membranes as well as meta-surface layouts.
	
	Finally, we demonstrated the use of \smuthi for the simulation of disordered volumetric aggregates of scattering nanoparticles in a dielectric micro film, showing the feasibility of simulations comprising \num{\sim e4} wavelength-scale particles in a layered medium.
	To the best of our knowledge, \smuthi is the only currently available tool that allows addressing this class of problems at such a large scale.
	
	We therefore believe that \smuthi will allow to expand the complexity of configurations that can be modeled rigorously, enabling the study of aspects that are traditionally difficult to investigate, such as finite-size effects, collective resonances, fabrication defects, effective medium approximations, scaling of transport properties, multi-scale heterogeneity and aperiodic structures in general.
	
	For the future, we plan two major additions to enhance the capability of \smuthi for the particularly interesting use case of photonic meta surfaces.
	These additions will address the run time, limited by the solution of the linear system \eqref{eq:linear system}, as well as the important case of close particles with intersecting circumscribing spheres.
	
	Finally, an extension of the simulation method to infinitely laterally periodic structures will be published in the near future.
	
	\section{Acknowledgments}
	We would like to thank the authors of the NFM-DS code (Adrian Doicu, Thomas Wriedt and Yuri Eremin) for allowing us to use their software in \smuthi.
	Furthermore, we thank H\r{a}kan Johansson for extending his pywigxjpf code \cite{johansson2016fast} to our needs.
	We acknowledge code additions and important user feedback from Giacomo Mazzamuto, Ilia Rasskazov, Fabio Mangini and Alan Zhan.
	LP acknowledges NVIDIA Corporation for the donation of a Titan Xp GPU through the GPU Grant program.
	KMC acknowledges support by the Polish National Science center via the project 2020/37/N/ST3/03334.
	DT acknowledges funding by the Deutsche Forschungsgemeinschaft (DFG, German Research Foundation) under Germany's
	Excellence Strategy via the Excellence Cluster 3D Matter Made to Order (EXC-2082/1 - 390761711) and the project GO 2615/2-1
	(project number 410400458) within the DFG-SPP 1839 ``Tailored Disorder''.
	KL and ASK acknowledge the support of the Russian Science Foundation (Project 21-72-30018).
	

\begin{thebibliography}{60}
\expandafter\ifx\csname natexlab\endcsname\relax\def\natexlab#1{#1}\fi
\providecommand{\bibinfo}[2]{#2}
\ifx\xfnm\relax \def\xfnm[#1]{\unskip,\space#1}\fi
\bibitem[{Jahani and Jacob(2016)}]{jahani2016all}
\bibinfo{author}{S.~Jahani}, \bibinfo{author}{Z.~Jacob},
\newblock \bibinfo{title}{All-dielectric metamaterials},
\newblock \bibinfo{journal}{Nature nanotechnology} \bibinfo{volume}{11}
  (\bibinfo{year}{2016}) \bibinfo{pages}{23--36}.
\bibitem[{Staude and Schilling(2017)}]{staude2017metamaterial}
\bibinfo{author}{I.~Staude}, \bibinfo{author}{J.~Schilling},
\newblock \bibinfo{title}{Metamaterial-inspired silicon nanophotonics},
\newblock \bibinfo{journal}{Nature photonics} \bibinfo{volume}{11}
  (\bibinfo{year}{2017}) \bibinfo{pages}{274--284}.
\bibitem[{Prieve(1999)}]{prieve1999measurement}
\bibinfo{author}{D.~C. Prieve},
\newblock \bibinfo{title}{Measurement of colloidal forces with {TIRM}},
\newblock \bibinfo{journal}{Advances in Colloid and Interface Science}
  \bibinfo{volume}{82} (\bibinfo{year}{1999}) \bibinfo{pages}{93--125}.
\bibitem[{Li et~al.(2010)Li, Huang, Ding, Yang, Li, Zhou, Fan, Zhang, Zhou, Ren
  et~al.}]{li2010shell}
\bibinfo{author}{J.~F. Li}, \bibinfo{author}{Y.~F. Huang},
  \bibinfo{author}{Y.~Ding}, \bibinfo{author}{Z.~L. Yang},
  \bibinfo{author}{S.~B. Li}, \bibinfo{author}{X.~S. Zhou},
  \bibinfo{author}{F.~R. Fan}, \bibinfo{author}{W.~Zhang},
  \bibinfo{author}{Z.~Y. Zhou}, \bibinfo{author}{B.~Ren}, et~al.,
\newblock \bibinfo{title}{Shell-isolated nanoparticle-enhanced {Raman}
  spectroscopy},
\newblock \bibinfo{journal}{Nature} \bibinfo{volume}{464}
  (\bibinfo{year}{2010}) \bibinfo{pages}{392--395}.
\bibitem[{Juan et~al.(2011)Juan, Righini, and Quidant}]{juan2011plasmon}
\bibinfo{author}{M.~L. Juan}, \bibinfo{author}{M.~Righini},
  \bibinfo{author}{R.~Quidant},
\newblock \bibinfo{title}{Plasmon nano-optical tweezers},
\newblock \bibinfo{journal}{Nature photonics} \bibinfo{volume}{5}
  (\bibinfo{year}{2011}) \bibinfo{pages}{349}.
\bibitem[{Saxena et~al.(2009)Saxena, Jain, and Mehta}]{saxena2009review}
\bibinfo{author}{K.~Saxena}, \bibinfo{author}{V.~Jain}, \bibinfo{author}{D.~S.
  Mehta},
\newblock \bibinfo{title}{A review on the light extraction techniques in
  organic electroluminescent devices},
\newblock \bibinfo{journal}{Optical Materials} \bibinfo{volume}{32}
  (\bibinfo{year}{2009}) \bibinfo{pages}{221--233}.
\bibitem[{Brongersma et~al.(2014)Brongersma, Cui, and
  Fan}]{brongersma2014light}
\bibinfo{author}{M.~L. Brongersma}, \bibinfo{author}{Y.~Cui},
  \bibinfo{author}{S.~Fan},
\newblock \bibinfo{title}{Light management for photovoltaics using high-index
  nanostructures},
\newblock \bibinfo{journal}{Nature materials} \bibinfo{volume}{13}
  (\bibinfo{year}{2014}) \bibinfo{pages}{451--460}.
\bibitem[{Yoshie et~al.(2004)Yoshie, Scherer, Hendrickson, Khitrova, Gibbs,
  Rupper, Ell, Shchekin, and Deppe}]{yoshie2004vacuum}
\bibinfo{author}{T.~Yoshie}, \bibinfo{author}{A.~Scherer},
  \bibinfo{author}{J.~Hendrickson}, \bibinfo{author}{G.~Khitrova},
  \bibinfo{author}{H.~Gibbs}, \bibinfo{author}{G.~Rupper},
  \bibinfo{author}{C.~Ell}, \bibinfo{author}{O.~Shchekin},
  \bibinfo{author}{D.~Deppe},
\newblock \bibinfo{title}{Vacuum {Rabi} splitting with a single quantum dot in
  a photonic crystal nanocavity},
\newblock \bibinfo{journal}{Nature} \bibinfo{volume}{432}
  (\bibinfo{year}{2004}) \bibinfo{pages}{200--203}.
\bibitem[{Kuznetsov et~al.(2016)Kuznetsov, Miroshnichenko, Brongersma, Kivshar,
  and Luk’yanchuk}]{kuznetsov2016optically}
\bibinfo{author}{A.~I. Kuznetsov}, \bibinfo{author}{A.~E. Miroshnichenko},
  \bibinfo{author}{M.~L. Brongersma}, \bibinfo{author}{Y.~S. Kivshar},
  \bibinfo{author}{B.~Luk’yanchuk},
\newblock \bibinfo{title}{Optically resonant dielectric nanostructures},
\newblock \bibinfo{journal}{Science} \bibinfo{volume}{354}
  (\bibinfo{year}{2016}).
\bibitem[{Babicheva and Evlyukhin(2017)}]{babicheva2017resonant}
\bibinfo{author}{V.~E. Babicheva}, \bibinfo{author}{A.~B. Evlyukhin},
\newblock \bibinfo{title}{Resonant lattice {Kerker} effect in metasurfaces with
  electric and magnetic optical responses},
\newblock \bibinfo{journal}{Laser \& Photonics Reviews} \bibinfo{volume}{11}
  (\bibinfo{year}{2017}) \bibinfo{pages}{1700132}.
\bibitem[{Riboli et~al.(2014)Riboli, Caselli, Vignolini, Intonti, Vynck,
  Barthelemy, Gerardino, Balet, Li, Fiore et~al.}]{riboli2014engineering}
\bibinfo{author}{F.~Riboli}, \bibinfo{author}{N.~Caselli},
  \bibinfo{author}{S.~Vignolini}, \bibinfo{author}{F.~Intonti},
  \bibinfo{author}{K.~Vynck}, \bibinfo{author}{P.~Barthelemy},
  \bibinfo{author}{A.~Gerardino}, \bibinfo{author}{L.~Balet},
  \bibinfo{author}{L.~H. Li}, \bibinfo{author}{A.~Fiore}, et~al.,
\newblock \bibinfo{title}{Engineering of light confinement in strongly
  scattering disordered media},
\newblock \bibinfo{journal}{Nature materials} \bibinfo{volume}{13}
  (\bibinfo{year}{2014}) \bibinfo{pages}{720--725}.
\bibitem[{Khanikaev et~al.(2013)Khanikaev, Mousavi, Tse, Kargarian, MacDonald,
  and Shvets}]{khanikaev2013photonic}
\bibinfo{author}{A.~B. Khanikaev}, \bibinfo{author}{S.~H. Mousavi},
  \bibinfo{author}{W.-K. Tse}, \bibinfo{author}{M.~Kargarian},
  \bibinfo{author}{A.~H. MacDonald}, \bibinfo{author}{G.~Shvets},
\newblock \bibinfo{title}{Photonic topological insulators},
\newblock \bibinfo{journal}{Nature materials} \bibinfo{volume}{12}
  (\bibinfo{year}{2013}) \bibinfo{pages}{233--239}.
\bibitem[{Wu and Hu(2015)}]{wu2015scheme}
\bibinfo{author}{L.-H. Wu}, \bibinfo{author}{X.~Hu},
\newblock \bibinfo{title}{Scheme for achieving a topological photonic crystal
  by using dielectric material},
\newblock \bibinfo{journal}{Physical Review Letters} \bibinfo{volume}{114}
  (\bibinfo{year}{2015}) \bibinfo{pages}{223901}.
\bibitem[{Hsu et~al.(2013)Hsu, Zhen, Lee, Chua, Johnson, Joannopoulos, and
  Solja{\v{c}}i{\'c}}]{hsu2013observation}
\bibinfo{author}{C.~W. Hsu}, \bibinfo{author}{B.~Zhen},
  \bibinfo{author}{J.~Lee}, \bibinfo{author}{S.-L. Chua},
  \bibinfo{author}{S.~G. Johnson}, \bibinfo{author}{J.~D. Joannopoulos},
  \bibinfo{author}{M.~Solja{\v{c}}i{\'c}},
\newblock \bibinfo{title}{Observation of trapped light within the radiation
  continuum},
\newblock \bibinfo{journal}{Nature} \bibinfo{volume}{499}
  (\bibinfo{year}{2013}) \bibinfo{pages}{188--191}.
\bibitem[{Waterman(1965)}]{waterman1965matrix}
\bibinfo{author}{P.~Waterman},
\newblock \bibinfo{title}{Matrix formulation of electromagnetic scattering},
\newblock \bibinfo{journal}{Proceedings of the IEEE} \bibinfo{volume}{53}
  (\bibinfo{year}{1965}) \bibinfo{pages}{805--812}.
\bibitem[{Peterson and Str{\"o}m(1973)}]{peterson1973t}
\bibinfo{author}{B.~Peterson}, \bibinfo{author}{S.~Str{\"o}m},
\newblock \bibinfo{title}{T-matrix for electromagnetic scattering from an
  arbitrary number of scatterers and representations of {E(3)}},
\newblock \bibinfo{journal}{Physical Review D} \bibinfo{volume}{8}
  (\bibinfo{year}{1973}) \bibinfo{pages}{3661}.
\bibitem[{Mackowski and Mishchenko(1996)}]{mackowski1996calculation}
\bibinfo{author}{D.~W. Mackowski}, \bibinfo{author}{M.~I. Mishchenko},
\newblock \bibinfo{title}{Calculation of the {T-matrix} and the scattering
  matrix for ensembles of spheres},
\newblock \bibinfo{journal}{JOSA A} \bibinfo{volume}{13} (\bibinfo{year}{1996})
  \bibinfo{pages}{2266--2278}.
\bibitem[{Mackowski and Mishchenko(2011)}]{mackowski2011multiple}
\bibinfo{author}{D.~Mackowski}, \bibinfo{author}{M.~Mishchenko},
\newblock \bibinfo{title}{A multiple sphere {T-matrix} {Fortran} code for use
  on parallel computer clusters},
\newblock \bibinfo{journal}{Journal of Quantitative Spectroscopy and Radiative
  Transfer} \bibinfo{volume}{112} (\bibinfo{year}{2011})
  \bibinfo{pages}{2182--2192}.
\bibitem[{Markkanen and Yuffa(2017)}]{markkanen2017fast}
\bibinfo{author}{J.~Markkanen}, \bibinfo{author}{A.~J. Yuffa},
\newblock \bibinfo{title}{Fast superposition {T-matrix} solution for clusters
  with arbitrarily-shaped constituent particles},
\newblock \bibinfo{journal}{Journal of Quantitative Spectroscopy and Radiative
  Transfer} \bibinfo{volume}{189} (\bibinfo{year}{2017})
  \bibinfo{pages}{181--188}.
\bibitem[{Egel et~al.(2017)Egel, Pattelli, Mazzamuto, Wiersma, and
  Lemmer}]{egel2017celes}
\bibinfo{author}{A.~Egel}, \bibinfo{author}{L.~Pattelli},
  \bibinfo{author}{G.~Mazzamuto}, \bibinfo{author}{D.~S. Wiersma},
  \bibinfo{author}{U.~Lemmer},
\newblock \bibinfo{title}{{CELES}: {CUDA}-accelerated simulation of
  electromagnetic scattering by large ensembles of spheres},
\newblock \bibinfo{journal}{Journal of Quantitative Spectroscopy and Radiative
  Transfer} \bibinfo{volume}{199} (\bibinfo{year}{2017})
  \bibinfo{pages}{103--110}.
\bibitem[{Kl{\"o}ckner et~al.(2012)Kl{\"o}ckner, Pinto, Lee, Catanzaro, Ivanov,
  and Fasih}]{klockner2012pycuda}
\bibinfo{author}{A.~Kl{\"o}ckner}, \bibinfo{author}{N.~Pinto},
  \bibinfo{author}{Y.~Lee}, \bibinfo{author}{B.~Catanzaro},
  \bibinfo{author}{P.~Ivanov}, \bibinfo{author}{A.~Fasih},
\newblock \bibinfo{title}{{PyCUDA} and {PyOpenCL}: A scripting-based approach
  to {GPU} run-time code generation},
\newblock \bibinfo{journal}{Parallel Computing} \bibinfo{volume}{38}
  (\bibinfo{year}{2012}) \bibinfo{pages}{157--174}.
\bibitem[{Mishchenko et~al.(2004)Mishchenko, Videen, Babenko, Khlebtsov, and
  Wriedt}]{mishchenko2004t}
\bibinfo{author}{M.~Mishchenko}, \bibinfo{author}{G.~Videen},
  \bibinfo{author}{V.~Babenko}, \bibinfo{author}{N.~Khlebtsov},
  \bibinfo{author}{T.~Wriedt},
\newblock \bibinfo{title}{{T-matrix} theory of electromagnetic scattering by
  particles and its applications: A comprehensive reference database},
\newblock \bibinfo{journal}{Journal of Quantitative Spectroscopy and Radiative
  Transfer} \bibinfo{volume}{88} (\bibinfo{year}{2004})
  \bibinfo{pages}{357--406}.
\bibitem[{Kristensson(1980)}]{kristensson1980electromagnetic}
\bibinfo{author}{G.~Kristensson},
\newblock \bibinfo{title}{Electromagnetic scattering from buried
  inhomogeneities -- a general three-dimensional formalism},
\newblock \bibinfo{journal}{Journal of Applied Physics} \bibinfo{volume}{51}
  (\bibinfo{year}{1980}) \bibinfo{pages}{3486--3500}.
\bibitem[{Mackowski(2008)}]{mackowski2008exact}
\bibinfo{author}{D.~W. Mackowski},
\newblock \bibinfo{title}{Exact solution for the scattering and absorption
  properties of sphere clusters on a plane surface},
\newblock \bibinfo{journal}{Journal of Quantitative Spectroscopy and Radiative
  Transfer} \bibinfo{volume}{109} (\bibinfo{year}{2008})
  \bibinfo{pages}{770--788}.
\bibitem[{Egel and Lemmer(2014)}]{egel2014dipole}
\bibinfo{author}{A.~Egel}, \bibinfo{author}{U.~Lemmer},
\newblock \bibinfo{title}{Dipole emission in stratified media with multiple
  spherical scatterers: Enhanced outcoupling from {OLEDs}},
\newblock \bibinfo{journal}{Journal of Quantitative Spectroscopy and Radiative
  Transfer} \bibinfo{volume}{148} (\bibinfo{year}{2014})
  \bibinfo{pages}{165--176}.
\bibitem[{Doicu and Wriedt(1999)}]{doicu1999calculation}
\bibinfo{author}{A.~Doicu}, \bibinfo{author}{T.~Wriedt},
\newblock \bibinfo{title}{Calculation of the {T-matrix} in the null-field
  method with discrete sources},
\newblock \bibinfo{journal}{JOSA A} \bibinfo{volume}{16} (\bibinfo{year}{1999})
  \bibinfo{pages}{2539--2544}.
\bibitem[{Doicu et~al.(2006)Doicu, Wriedt, and Eremin}]{doicu2006light}
\bibinfo{author}{A.~Doicu}, \bibinfo{author}{T.~Wriedt}, \bibinfo{author}{Y.~A.
  Eremin}, \bibinfo{title}{Light scattering by systems of particles}, volume
  \bibinfo{volume}{124}, \bibinfo{publisher}{Springer}, \bibinfo{year}{2006}.
\bibitem[{Egel et~al.(2017)Egel, Gomard, Kettlitz, and
  Lemmer}]{egel2017accurate}
\bibinfo{author}{A.~Egel}, \bibinfo{author}{G.~Gomard}, \bibinfo{author}{S.~W.
  Kettlitz}, \bibinfo{author}{U.~Lemmer},
\newblock \bibinfo{title}{Accurate optical simulation of nano-particle based
  internal scattering layers for light outcoupling from organic light emitting
  diodes},
\newblock \bibinfo{journal}{Journal of Optics} \bibinfo{volume}{19}
  (\bibinfo{year}{2017}) \bibinfo{pages}{025605}.
\bibitem[{Stein(1961)}]{stein1961addition}
\bibinfo{author}{S.~Stein},
\newblock \bibinfo{title}{Addition theorems for spherical wave functions},
\newblock \bibinfo{journal}{Quarterly of Applied Mathematics}
  \bibinfo{volume}{19} (\bibinfo{year}{1961}) \bibinfo{pages}{15--24}.
\bibitem[{Cruzan(1962)}]{cruzan1962translational}
\bibinfo{author}{O.~R. Cruzan},
\newblock \bibinfo{title}{Translational addition theorems for spherical vector
  wave functions},
\newblock \bibinfo{journal}{Quarterly of Applied Mathematics}
  \bibinfo{volume}{20} (\bibinfo{year}{1962}) \bibinfo{pages}{33--40}.
\bibitem[{Mishchenko et~al.(2002)Mishchenko, Travis, and
  Lacis}]{mishchenko2002scattering}
\bibinfo{author}{M.~I. Mishchenko}, \bibinfo{author}{L.~D. Travis},
  \bibinfo{author}{A.~A. Lacis}, \bibinfo{title}{Scattering, absorption, and
  emission of light by small particles}, \bibinfo{publisher}{Cambridge
  University Press}, \bibinfo{year}{2002}.
\bibitem[{Johansson and Forss\'en(2016)}]{johansson2016fast}
\bibinfo{author}{H.~T. Johansson}, \bibinfo{author}{C.~Forss\'en},
\newblock \bibinfo{title}{Fast and accurate evaluation of {Wigner} $3j$, $6j$,
  and $9j$ symbols using prime factorization and multiword integer arithmetic},
\newblock \bibinfo{journal}{SIAM Journal on Scientific Computing}
  \bibinfo{volume}{38} (\bibinfo{year}{2016}) \bibinfo{pages}{A376--A384}.
\bibitem[{Egel(2019)}]{egel2019accurate}
\bibinfo{author}{A.~Egel}, \bibinfo{title}{Accurate optical simulation of
  disordered scattering layers for light extraction from organic light emitting
  diodes}, Ph.D. thesis, Karlsruhe Institue of Technology,
  \bibinfo{year}{2019}.
\bibitem[{Colab(2021)}]{colab}
Colab, \bibinfo{title}{Google {Colaboratory}},
  \bibinfo{howpublished}{\url{https://colab.research.google.com}},
  \bibinfo{year}{2021}.
\bibitem[{Geuzaine and Remacle(2009)}]{geuzaine2009gmsh}
\bibinfo{author}{C.~Geuzaine}, \bibinfo{author}{J.-F. Remacle},
\newblock \bibinfo{title}{Gmsh: A {3-D} finite element mesh generator with
  built-in pre- and post-processing facilities},
\newblock \bibinfo{journal}{International journal for numerical methods in
  engineering} \bibinfo{volume}{79} (\bibinfo{year}{2009})
  \bibinfo{pages}{1309--1331}.
\bibitem[{{Dawson-Haggerty et al.}(2021)}]{trimesh}
\bibinfo{author}{{Dawson-Haggerty et al.}}, \bibinfo{title}{trimesh},
  \bibinfo{howpublished}{\url{https://trimsh.org/}}, \bibinfo{year}{2021}.
\bibitem[{Egel et~al.(2016)Egel, Theobald, Donie, Lemmer, and
  Gomard}]{egel2016light}
\bibinfo{author}{A.~Egel}, \bibinfo{author}{D.~Theobald},
  \bibinfo{author}{Y.~Donie}, \bibinfo{author}{U.~Lemmer},
  \bibinfo{author}{G.~Gomard},
\newblock \bibinfo{title}{Light scattering by oblate particles near planar
  interfaces: on the validity of the {T-matrix} approach},
\newblock \bibinfo{journal}{Optics Express} \bibinfo{volume}{24}
  (\bibinfo{year}{2016}) \bibinfo{pages}{25154--25168}.
\bibitem[{Egel et~al.(2017)Egel, Eremin, Wriedt, Theobald, Lemmer, and
  Gomard}]{egel2017extending}
\bibinfo{author}{A.~Egel}, \bibinfo{author}{Y.~Eremin},
  \bibinfo{author}{T.~Wriedt}, \bibinfo{author}{D.~Theobald},
  \bibinfo{author}{U.~Lemmer}, \bibinfo{author}{G.~Gomard},
\newblock \bibinfo{title}{Extending the applicability of the {T-matrix} method
  to light scattering by flat particles on a substrate via truncation of
  {Sommerfeld} integrals},
\newblock \bibinfo{journal}{Journal of Quantitative Spectroscopy and Radiative
  Transfer} \bibinfo{volume}{202} (\bibinfo{year}{2017})
  \bibinfo{pages}{279--285}.
\bibitem[{Theobald et~al.(2017)Theobald, Egel, Gomard, and
  Lemmer}]{theobald2017plane}
\bibinfo{author}{D.~Theobald}, \bibinfo{author}{A.~Egel},
  \bibinfo{author}{G.~Gomard}, \bibinfo{author}{U.~Lemmer},
\newblock \bibinfo{title}{Plane-wave coupling formalism for {T-matrix}
  simulations of light scattering by nonspherical particles},
\newblock \bibinfo{journal}{Physical Review A} \bibinfo{volume}{96}
  (\bibinfo{year}{2017}) \bibinfo{pages}{033822}.
\bibitem[{Purcell(1995)}]{purcell1995spontaneous}
\bibinfo{author}{E.~M. Purcell},
\newblock \bibinfo{title}{Spontaneous emission probabilities at radio
  frequencies},
\newblock in: \bibinfo{booktitle}{Confined Electrons and Photons},
  \bibinfo{publisher}{Springer}, \bibinfo{year}{1995}, pp.
  \bibinfo{pages}{839--839}.
\bibitem[{Neves and Pisignano(2012)}]{neves2012effect}
\bibinfo{author}{A.~A.~R. Neves}, \bibinfo{author}{D.~Pisignano},
\newblock \bibinfo{title}{Effect of finite terms on the truncation error of
  {Mie} series},
\newblock \bibinfo{journal}{Optics letters} \bibinfo{volume}{37}
  (\bibinfo{year}{2012}) \bibinfo{pages}{2418--2420}.
\bibitem[{Wiscombe(1980)}]{wiscombe1980improved}
\bibinfo{author}{W.~J. Wiscombe},
\newblock \bibinfo{title}{Improved {Mie} scattering algorithms},
\newblock \bibinfo{journal}{Applied optics} \bibinfo{volume}{19}
  (\bibinfo{year}{1980}) \bibinfo{pages}{1505--1509}.
\bibitem[{Chew(1995)}]{chew1995waves}
\bibinfo{author}{W.~C. Chew}, \bibinfo{title}{Waves and fields in inhomogeneous
  media}, \bibinfo{publisher}{IEEE press}, \bibinfo{year}{1995}.
\bibitem[{Egel et~al.(2016)Egel, Kettlitz, and Lemmer}]{egel2016efficient}
\bibinfo{author}{A.~Egel}, \bibinfo{author}{S.~W. Kettlitz},
  \bibinfo{author}{U.~Lemmer},
\newblock \bibinfo{title}{Efficient evaluation of {Sommerfeld} integrals for
  the optical simulation of many scattering particles in planarly layered
  media},
\newblock \bibinfo{journal}{JOSA A} \bibinfo{volume}{33} (\bibinfo{year}{2016})
  \bibinfo{pages}{698--706}.
\bibitem[{Sinev et~al.(2016)Sinev, Iorsh, Bogdanov, Permyakov, Komissarenko,
  Mukhin, Samusev, Valuckas, Kuznetsov, Luk'yanchuk
  et~al.}]{sinev2016polarization}
\bibinfo{author}{I.~Sinev}, \bibinfo{author}{I.~Iorsh},
  \bibinfo{author}{A.~Bogdanov}, \bibinfo{author}{D.~Permyakov},
  \bibinfo{author}{F.~Komissarenko}, \bibinfo{author}{I.~Mukhin},
  \bibinfo{author}{A.~Samusev}, \bibinfo{author}{V.~Valuckas},
  \bibinfo{author}{A.~I. Kuznetsov}, \bibinfo{author}{B.~S. Luk'yanchuk},
  et~al.,
\newblock \bibinfo{title}{Polarization control over electric and magnetic
  dipole resonances of dielectric nanoparticles on metallic films},
\newblock \bibinfo{journal}{Laser \& Photonics Reviews} \bibinfo{volume}{10}
  (\bibinfo{year}{2016}) \bibinfo{pages}{799--806}.
\bibitem[{Martin(2019)}]{martin2019t}
\bibinfo{author}{T.~Martin},
\newblock \bibinfo{title}{{T-matrix} method for closely adjacent obstacles},
\newblock \bibinfo{journal}{Journal of Quantitative Spectroscopy and Radiative
  Transfer} \bibinfo{volume}{234} (\bibinfo{year}{2019})
  \bibinfo{pages}{40--46}.
\bibitem[{Lumerical(2021)}]{lumerical}
Lumerical, \bibinfo{title}{Lumerical {Inc.}},
  \bibinfo{howpublished}{\url{https://www.lumerical.com/}},
  \bibinfo{year}{2021}.
\bibitem[{Akahane et~al.(2003)Akahane, Asano, Song, and Noda}]{akahane2003high}
\bibinfo{author}{Y.~Akahane}, \bibinfo{author}{T.~Asano},
  \bibinfo{author}{B.-S. Song}, \bibinfo{author}{S.~Noda},
\newblock \bibinfo{title}{High-{Q} photonic nanocavity in a two-dimensional
  photonic crystal},
\newblock \bibinfo{journal}{Nature} \bibinfo{volume}{425}
  (\bibinfo{year}{2003}) \bibinfo{pages}{944--947}.
\bibitem[{Ding et~al.(2016)Ding, Yi, Li, Ren, Wu, Panneerselvam, and
  Tian}]{ding2016nanostructure}
\bibinfo{author}{S.-Y. Ding}, \bibinfo{author}{J.~Yi}, \bibinfo{author}{J.-F.
  Li}, \bibinfo{author}{B.~Ren}, \bibinfo{author}{D.-Y. Wu},
  \bibinfo{author}{R.~Panneerselvam}, \bibinfo{author}{Z.-Q. Tian},
\newblock \bibinfo{title}{Nanostructure-based plasmon-enhanced {Raman}
  spectroscopy for surface analysis of materials},
\newblock \bibinfo{journal}{Nature Reviews Materials} \bibinfo{volume}{1}
  (\bibinfo{year}{2016}) \bibinfo{pages}{1--16}.
\bibitem[{COMSOL(2021)}]{comsol}
COMSOL, \bibinfo{title}{Comsol {Multiphysics}},
  \bibinfo{howpublished}{\url{https://www.comsol.com/}}, \bibinfo{year}{2021}.
\bibitem[{Li(1980)}]{li1980refractive}
\bibinfo{author}{H.~Li},
\newblock \bibinfo{title}{Refractive index of silicon and germanium and its
  wavelength and temperature derivatives},
\newblock \bibinfo{journal}{Journal of Physical and Chemical Reference Data}
  \bibinfo{volume}{9} (\bibinfo{year}{1980}) \bibinfo{pages}{561--658}.
\bibitem[{De~Sturler(1999)}]{de1999truncation}
\bibinfo{author}{E.~De~Sturler},
\newblock \bibinfo{title}{Truncation strategies for optimal {Krylov} subspace
  methods},
\newblock \bibinfo{journal}{SIAM Journal on Numerical Analysis}
  \bibinfo{volume}{36} (\bibinfo{year}{1999}) \bibinfo{pages}{864--889}.
\bibitem[{Theobald et~al.(2020)Theobald, Yu, Gomard, and
  Lemmer}]{theobald2020design}
\bibinfo{author}{D.~Theobald}, \bibinfo{author}{S.~Yu},
  \bibinfo{author}{G.~Gomard}, \bibinfo{author}{U.~Lemmer},
\newblock \bibinfo{title}{Design of selective reflectors utilizing multiple
  scattering by core--shell nanoparticles for color conversion films},
\newblock \bibinfo{journal}{ACS Photonics} \bibinfo{volume}{7}
  (\bibinfo{year}{2020}) \bibinfo{pages}{1452--1460}.
\bibitem[{Pidgayko et~al.(2020)Pidgayko, Sadrieva, Ladutenko, and
  Bogdanov}]{pidgayko2020polarization}
\bibinfo{author}{D.~Pidgayko}, \bibinfo{author}{Z.~Sadrieva},
  \bibinfo{author}{K.~Ladutenko}, \bibinfo{author}{A.~Bogdanov},
\newblock \bibinfo{title}{Polarization-controlled selective excitation of {Mie}
  resonances in a dielectric nanoparticle on a coated substrate},
\newblock \bibinfo{journal}{Physical Review B} \bibinfo{volume}{102}
  (\bibinfo{year}{2020}) \bibinfo{pages}{245406}.
\bibitem[{Dvoretckaia et~al.(2020)Dvoretckaia, Ladutenko, Mozharov, Zograf,
  Bogdanov, and Mukhin}]{dvoretckaia2020electrically}
\bibinfo{author}{L.~Dvoretckaia}, \bibinfo{author}{K.~Ladutenko},
  \bibinfo{author}{A.~Mozharov}, \bibinfo{author}{G.~Zograf},
  \bibinfo{author}{A.~Bogdanov}, \bibinfo{author}{I.~Mukhin},
\newblock \bibinfo{title}{Electrically driven metal and all-dielectric
  nanoantennas for plasmon polariton excitation},
\newblock \bibinfo{journal}{Journal of Quantitative Spectroscopy and Radiative
  Transfer} \bibinfo{volume}{244} (\bibinfo{year}{2020})
  \bibinfo{pages}{106825}.
\bibitem[{Czajkowski and Antosiewicz(2019)}]{czajkowski2019electromagnetic}
\bibinfo{author}{K.~M. Czajkowski}, \bibinfo{author}{T.~J. Antosiewicz},
\newblock \bibinfo{title}{Electromagnetic coupling in optical devices based on
  random arrays of dielectric nanoresonators},
\newblock \bibinfo{journal}{The Journal of Physical Chemistry C}
  \bibinfo{volume}{124} (\bibinfo{year}{2019}) \bibinfo{pages}{896--905}.
\bibitem[{Czajkowski et~al.(2020)Czajkowski, Bancerek, and
  Antosiewicz}]{czajkowski2020multipole}
\bibinfo{author}{K.~M. Czajkowski}, \bibinfo{author}{M.~Bancerek},
  \bibinfo{author}{T.~J. Antosiewicz},
\newblock \bibinfo{title}{Multipole analysis of substrate-supported dielectric
  nanoresonator metasurfaces via the {T-matrix} method},
\newblock \bibinfo{journal}{Physical Review B} \bibinfo{volume}{102}
  (\bibinfo{year}{2020}) \bibinfo{pages}{085431}.
\bibitem[{Warren et~al.(2020)Warren, Alkaisi, and Moore}]{warren2020design}
\bibinfo{author}{A.~Warren}, \bibinfo{author}{M.~Alkaisi},
  \bibinfo{author}{C.~Moore},
\newblock \bibinfo{title}{Design of {2D} plasmonic diffraction gratings for
  sensing and super-resolution imaging applications},
\newblock in: \bibinfo{booktitle}{2020 IEEE International Instrumentation and
  Measurement Technology Conference (I2MTC)}, \bibinfo{organization}{IEEE}, pp.
  \bibinfo{pages}{1--6}.
\bibitem[{Boscolo and Midrio(2004)}]{boscolo2004three}
\bibinfo{author}{S.~Boscolo}, \bibinfo{author}{M.~Midrio},
\newblock \bibinfo{title}{Three-dimensional multiple-scattering technique for
  the analysis of photonic-crystal slabs},
\newblock \bibinfo{journal}{Journal of lightwave technology}
  \bibinfo{volume}{22} (\bibinfo{year}{2004}) \bibinfo{pages}{2778}.
\bibitem[{Pissoort et~al.(2007)Pissoort, Michielssen, Ginste, and
  Olyslager}]{pissoort2007fast}
\bibinfo{author}{D.~Pissoort}, \bibinfo{author}{E.~Michielssen},
  \bibinfo{author}{D.~V. Ginste}, \bibinfo{author}{F.~Olyslager},
\newblock \bibinfo{title}{Fast-multipole analysis of electromagnetic scattering
  by photonic crystal slabs},
\newblock \bibinfo{journal}{Journal of lightwave technology}
  \bibinfo{volume}{25} (\bibinfo{year}{2007}) \bibinfo{pages}{2847--2863}.

\end{thebibliography}
	
	

\end{document}